\documentclass[preprint,tightenlines,aps,floats,nofootinbib,amssymb,groupedaddress]{revtex4-1}
\pdfoutput=1
\usepackage[sort&compress]{natbib}
\usepackage{epsf,epsfig}
\usepackage{amsmath}
\usepackage{hyperref}
\usepackage{subfigure}
\usepackage{graphicx}
\usepackage{color}
\usepackage{mathrsfs}

\newcommand{\be}{\begin{equation}}
\newcommand{\ee}{\end{equation}}
\newcommand{\ba}{\begin{array}}
\newcommand{\eaq}{\end{array}}
\newcommand{\bea}{\begin{eqnarray}}
\newcommand{\eea}{\end{eqnarray}}

\newcommand{\nn}{\nonumber}

\newcommand{\bi}{\begin{itemize}}
\newcommand{\ei}{\end{itemize}}
\newcommand{\bal}{\begin{aligned}}
\newcommand{\eal}{\end{aligned}}
\newcommand{\Tr}{\operatorname{Tr}}
\newcommand{\Exp}{\operatorname{Exp}}

\usepackage{comment}
\makeatletter

\begin{document}

\title{Holographic description of $SO(5) \rightarrow SO(4)$ composite Higgs model}
\author{D. Espriu}
\author{A. Katanaeva}
\affiliation{\it Departament de F\'\i sica Qu\`antica i Astrof\'\i sica and \\
Institut de Ci\`encies del Cosmos (ICCUB), Universitat de Barcelona,\\ 
Mart\'i i Franqu\`es 1, 08028 Barcelona, Catalonia, Spain}

\vspace*{3cm}

\thispagestyle{empty}

\begin{abstract}
We study a 5D bottom-up  holographic model that is expected to describe the dynamics of the minimal 
composite Higgs model characterized by a $SO(5)\to SO(4)$ global symmetry breaking pattern. 
We assume that the fundamental degrees of freedom are scalars transforming under some 
representation of $SO(5)$ and subject to some
unspecified strong interactions. The holographic description presented here
is inspired by previous studies performed in the context of QCD and it allows for the consideration
of spin one and spin zero resonances. The resulting spectrum leads in a natural way to a variety of
resonances. Namely, those transforming under the unbroken $SO(4)$ subgroup exhibit an exact 
degeneracy between the two Regge trajectories of vector and scalar channels, 
while the resonances with quantum numbers 
in the $SO(5)/ SO(4)$ coset lie on a trajectory of (heavier) spin one states and a non-degenerated one
of scalar resonances where the four lowest lying ones are massless. 
These correspond to the four Goldstone bosons in 4D associated to the 
global symmetry breaking pattern. Restrictions derived from the experimental constraints (Higgs couplings, 
$S$ parameter, etc.) are then implemented and we conclude that the model is able to accommodate 
vector and scalar resonances with masses in the range
1 TeV to 2 TeV without encountering phenomenological difficulties. Extension to generic models 
characterized by the breaking pattern $SO(N)\to SO(N^\prime<N)$ is straightforward.

\end{abstract}

\maketitle

\section{Introduction}
To this date the fundamental nature of the electroweak symmetry breaking sector (EWSBS) of
the Standard Model (SM) has not been fully elucidated yet. The majority of the data available
so far seems to indicate that the minimal version of the SM with a doublet of complex scalar fields
is fully compatible with the data.

However, the jury is still out. There are models where possible departures with respect to the 
predictions of the SM are small in a natural way. 
Misaligned composite Higgs models~\cite{kaplan84-1, *kaplan84-2, *kaplan84-3, *kaplan84-4, *kaplan85}
could be considered as a paradigm of this type of theories. In these models a global 
symmetry group $\mathcal G$ is broken down to a subgroup $\mathcal H$, 
that contains the SM global group $SU(2)\times U(1)$. The SM gauge group itself $\mathcal H^\prime$ is rotated 
with respect to $\mathcal H$ by a certain angle $\theta$ around one of the broken directions. 
The value of $\theta$ is determined 
dynamically and it is, roughly speaking, determined by a contribution from the top quark to the 
Higgs potential
(that happens to break the custodial symmetry represented by the global $SU(2)\times SU(2)$ (sub)group
of $\mathcal H$).

In these models it is a priori possible to establish a large hierarchy between the scale at which the global 
symmetry breaks (say $4\pi F$), presumably due to some QCD-like new strong interactions, and the 
weak scale characterized by the Fermi scale $v$, related to the one
where in Higgsless models weak interactions may become non-perturbative ($4\pi v$). The latter is
customarily regarded as the scale where an extended EWSBS can become strongly interacting.
However, a large scale separation, i.e. $F\gg v$  may lead
to a relevant  amount of fine-tuning in order to keep light the states that should remain in the low
energy part of the spectrum.

Having said that, it is a fact that not much is known about the dynamics and the spectrum of theories
such as the ones just described. This is particularly true for the models
where the global symmetry $\mathcal G$ cannot be
realized with fermions at the microscopic level, 
among those is the one introduced first in~\cite{ACP_2005} with
$\mathcal G=SO(5)$ and $\mathcal H=SO(4)\simeq SU(2) \times SU(2)$, called the minimal one
for providing the most economical way to preserve the custodial symmetry. Yet it is often implicitly assumed 
that the spectrum in
such models can be inferred from what we have learned from QCD, the only relativistic 
strongly interacting theory that we are familiar with. 

It is known that one can get a fairly accurate description of QCD using the so-called bottom-up holographic 
models, where space-time is extended with an additional dimension $z$, and assumed to be described by an 
anti-de Sitter (AdS) metric. The value $z=0$ corresponds to the UV brane, where the theory is assumed 
to be described by a conformal field theory (CFT) 
as befits a critical point of QCD at short distances. In the IR the holographic model
should reproduce the fact that QCD breaks conformality becoming a confining theory.
There are two general approaches to this issue. The first one is to introduce an infrared brane, i.e. 
to restrict the metric of a model to be a slice of the AdS metric; this is
the hard wall (HW) proposal \cite{HW_2005, daRold_2005}. The second way is to make the AdS metric smoothly cut-off
at large $z$; it was put forward in \cite{SW_2006} and referred to as 
 the soft wall (SW) model. 
Originally inspired by formal developments that establish an exact correspondence
between the string theories on $AdS_5\times S_5$  and $\mathcal N=4$ super Yang-Mills
gauge theory on $\partial AdS_5$ \cite{Maldacena_1999,*Gubser1998,*Witten_1998}, the bottom-up 
holographic models are just conjectural. Nevertheless, they 
provide a surprisingly accurate description of several facets of QCD, 
such as chiral symmetry breaking in HW and the phenomenological spectra in SW. 
It is therefore appropriate to try and use these techniques to get precious information on
theories whose dynamics is not really known.

The holographic approach has been used before to give a new insight to the known 
mechanisms of the electroweak symmetry breaking.  
The minimal composite Higgs scenario was realized first in
\cite{ACP_2005, AC_2005} using a HW model, even though the first example of the technique was proposed
for the simplest case of the $SU(3)\rightarrow SU(2)$ breaking pattern in~\cite{Contino_Nomura_2003}.
These models have the following characteristics. The gauge symmetry of
the SM is extended to the bulk and the symmetry breaking pattern totally relies on 
the two branes being introduced; the boundary conditions for the 5D fields determine whether 
they correspond to the dynamical fields or not. 
The Higgs is associated with the fifth component of gauge fields in the direction of the broken gauge symmetry.
The Higgs  potential  is absent at the tree-level and is  determined  by  quantum
corrections (dominantly gauge bosons and top quarks at one-loop level).  The extensive study in \cite{ACP_2005, AC_2005}
includes a complete calculation of the Higgs potential and analysis of several electroweak
observables ($S,\ T,\ Z\rightarrow b\overline{b}$). The stress is made on a way one embeds
SM quarks into 5D model (choosing representation etc.) as their contribution is crucial for
most of the mentioned computations.

In the present study we use a SW model approach and lay emphasis on an alternative way
to realize the global symmetry breaking pattern and to introduce spin zero fields.
The $SO(5) \rightarrow SO(4)$ breaking happens in the bulk Lagrangian of the scalar fields,
reminiscent to the one of the generalized sigma model used for QCD~\cite{gasiorowicz, *Fariborz,
*Rischke2010}.
The Goldstone bosons are introduced explicitly and there is no gauge-Higgs unification
characteristic to the former studies in the SW framework~\cite{Falkowski2008}. Gauging of the SM may be
 achieved via extending
the 5D covariant derivative with electroweak bosons, but they are assumed 
to have no $z$ propagation. They are treated perturbatively 
on the UV brane and holography is only really relevant to determine the 
corresponding correlation functions.
It is important to emphasize that this approach is quite different from the one adopted in \cite{ACP_2005, AC_2005}. 
In the present proposal the dynamics responsible for the $SO(5)\to SO(4)$ breaking is entirely `decoupled' 
from the SM gauge fields. The latter are treated in fact as external sources that do not 
participate in the strong dynamics (except eventually through mixing of fields
with identical quantum numbers). We do not consider SM fermion fields either, which in composite Higgs
scenarios are essential to provide the Higgs potential~\cite{Bellazzini2014,Panico2016} giving 
the Higgs mass and self-couplings, among other things. We adopt
the point of view that the said potential is of perturbative origin and holographic techniques are not applicable.
Needless to say that no new insight into the naturalness problem nor the 
origin of the hierarchy of the various scales involved is provided. We just attempt to describe the strong 
dynamics behind the composite sector, the resulting 
spectrum and verify the fulfilment of the expected current algebra properties, such as the Weinberg sum rules, together with
the existing constraints from electroweak precision measurements. In a separate work we will describe the model predictions
on coupling constants and form factors of the various resonances resulting from the strong dynamics. Taken together these 
investigations could shed light on the existence or not of strong dynamics associated to the EWSBS.

Our specific treatment is inspired by various
bottom-up holographic approaches to QCD \cite{HW_2005, daRold_2005, SW_2006, Hirn2005,*Hirn2006, DAmbrosio2015} but the 
spectrum and several properties are quite
distinct from them as we will see. 
We have the necessity to define a fundamental theory (made out of scalar fields in the present case) 
in order to determine the scalar operators and the conserved currents
and to match the normalizations of spin zero and spin one sectors. Other possibilities 
for the fundamental degrees of freedom are conceivable, but 
the one presented here is the simplest one.

\section{General construction}
\subsection{Strongly interacting sector: operators and misalignment}
We will consider extending the SM with an additional strongly interacting sector, 
assumed to be conformal in the UV. 
This sector is endowed with a global symmetry that is spontaneously broken.
The symmetry breaking pattern is $\mathcal G \rightarrow \mathcal H$
and the coset $\mathcal G / \mathcal H$ contains a representation of the Goldstone bosons corresponding to 
the quantum numbers of the Higgs doublet, and possibly other Goldstone bosons depending on the groups.
The coupling to the electroweak sector of the SM is implemented through
the conserved currents of the new sector
\be\label{lagr1}
\mathcal L = \mathcal {\widetilde L}_{str. int.} + \mathcal L_{SM} + \widetilde J^{a_L\ \mu}  W_\mu^{a_L}
+ \widetilde J^{Y\ \mu}  B_\mu.
\ee
The currents of the
strongly interacting sector $J^{a_L\ \mu}$ and $J^{Y\ \mu}$ contain the generators
of the $SU(2)_L\times U(1)$  
global group that is necessarily included in $\mathcal H$.  Moreover, these generators 
belonging to $\mathcal H$ are rotated (and marked with tildes)
with respect to the SM gauge group $\mathcal H'$ containing the $W$ and $B$ electroweak
gauge bosons. Similarly, we mark with a tilde $\mathcal {L}_{str. int.}$, corresponding to the
strongly interacting sector. We will describe this point in more detail below. 

Let us now concentrate on the minimal composite Higgs model (MCHM), where the global symmetry breaking pattern 
is realized as $SO(5) \rightarrow SO(4)$.
We denote by $T^A, \ A=1,...,10$ the generators of $SO(5)$, 
represented standardly by $5\times5$ matrices, which are traceless $\Tr T^A=0$ and unless otherwise stated have
the established normalization $\Tr (T^AT^B)=\delta^{AB}$. 
They separate naturally into two groups:
\bi
\item the unbroken generators, in the case of MCHM those of $SO(4)\backsimeq SU(2)_L\times SU(2)_R$;
$T^a, \ a=1,...,6$ :
\be \label{genso4}
    T^\alpha_L=\begin{pmatrix}
    t^\alpha_L & 0\\
    0 & 0\\
   \end{pmatrix},\ T^\alpha_R=\begin{pmatrix}
    t^\alpha_R & 0\\
    0 & 0\\
   \end{pmatrix}, \ \alpha=1,2,3,
\ee
where $(t^\alpha_{L/R})_{jk}=-\frac i2 (\varepsilon_{\alpha\beta\gamma}\delta^\beta_j\delta^\gamma_k
\pm (\delta_j^\alpha\delta_k^4-\delta_k^\alpha\delta_j^4))$, $j,k=1,...,4$.
\item the broken generators, corresponding to the coset $SO(5)/SO(4)$; $\widehat{T}^i,\  i=1,2,3,4$ : 
\be\widehat{T}^i_{IJ}=-\frac i{\sqrt 2}(\delta^i_I\delta^5_J-\delta^i_J\delta^5_I),\ \ I,J=1,...,5. \ee

\ei

The MCHM does not admit complex Dirac fermions as fundamental fields at the microscopic level due to the nature
of the global symmetry group.
Let us assume instead that the theory contains some fundamental scalars transforming under the 
global $SO(5)$ symmetry. We choose  a rank 2 tensor representation, so a fundamental field is a general 
$5\times5$ matrix $s^{\alpha\beta}$. The Lagrangian invariant under the global 
$s\rightarrow gs g^{-1}, \ g\in SO(5)$ transformation
is $\mathcal L =\frac12 \partial_\mu s_{\alpha\beta}\partial^\mu s^\top_{\beta\alpha}
-\frac12 m^2 s_{\alpha\beta}s^\top_{\beta\alpha}$
. We can construct a scalar invariant $s^{\alpha\gamma}s^{\gamma\alpha}$, giving a scalar operator 
$\mathcal O_S^{\alpha\beta}(x)=s^{\alpha\gamma}s^{\gamma\beta}$ with dimension $\Delta=2$, spin $p=0$; and a Noether current 
$i[T^A,s]_{\alpha\beta}\partial^\mu s^\top_{\beta\alpha}$ giving a vector operator $\mathcal O_V^{A\ \mu}(x)$, with $\Delta=3$, $p=1$.
These define the currents of Eqn.~(\ref{lagr1}):
\bi
\item for $A=a_L$: $J^{a_L}_\mu=\frac{g}{\sqrt2}\mathcal O^{a_L}_\mu(x)$;
\item for $A=3_R$: $J^Y_{\mu}=\frac{g'}{\sqrt2}\mathcal O^{3_R}_\mu(x)$, the hypercharge is assumed to 
be realized  as $Y=T_{3_R}$.
\ei
The coupling coefficients are chosen to be in concordance with the usual SM normalization 
of the electroweak generators in the resulting expressions.

Having determined the symmetry breaking pattern in the strongly interacting sector we may come back to
the discussion of the vacuum misalignment phenomenon leading to 
the EWSB. A quantity parametrizing this breaking is a rotation angle $\theta$ that relates
 the linearly-realized global group $\mathcal H=SO(4)$
and a gauged  group $\mathcal H'=SO(4)'$ containing the electroweak bosons in its subgroup $SU(2)'_L\times U(1)'$.
 It is natural to denote the generators of $SO(5) \rightarrow SO(4)'$ as $\{T^a(0),\ \widehat{T}^i(0)\}$
 and of $SO(5) \rightarrow SO(4)$ as $\{T^a(\theta),\ \widehat{T}^i(\theta)\}$ so that $\theta=0$ is assigned to the SM.
 
 We may choose any direction as the one preferred by the $SO(4)'$ and then make the misalignment occur with respect to it, this
 leads to a connection between the generators such as
\be
T^\alpha(\theta)=r(\theta)T^\alpha(0)r^{-1}(\theta), \ \text{with} \ r(\theta)=\begin{pmatrix}
          1_{3\times3} & 0 & 0\\
          0 &\cos(\theta)&\sin(\theta)\\
          0 & -\sin(\theta)&\cos(\theta)\\
         \end{pmatrix}.
\ee

This only affects the relation between the SM gauge generators and those of the strongly interacting sector, hence
the tildes in Eqn.~(\ref{lagr1}). 
A 5D dual model we proceed to define in the next subsection is not influenced by the misalignment effect.
It is an alternative way to describe the physics involved in the  $SO(5) \rightarrow SO(4)$ symmetry breaking,
derived from some interactions becoming strong in the infrared.

\subsection{5D model Lagrangian}
\label{sec-model}
In this subsection we describe the holographic 5D model realizing some conceptual features of the 4D MCHM.
See, for instance, the review~\cite{Panico2016} presenting various aspects of 4D studies, 
though we do not address at all the flavour issues discussed in abundance in the literature.

We have selected two composite operators -- a vector and a scalar one, to be defining to the theory, and hence
we have spin one and spin zero fields at the 5D side. The choice is motivated by the AdS/QCD models where a quark
bilinear and left- and right-handed currents corresponding to the chiral flavour symmetry are supposed to be important
for the description of the chiral dynamics.

The 5D AdS metric (with the radius $R$) is given by
\be
g_{MN}dx^M dx^N=\frac{R^2}{z^2}(\eta_{\mu\nu}dx^\mu dx^\nu-d^2z),\quad \eta_{\mu\nu}=\text{diag}(1,-1,-1,-1).
\ee
The $SO(5)$ invariant action that we assume has the following form 
\begin{align}\label{5Daction}
&S_{5D}=-\frac{1}{4g_5^2}\int d^5x\sqrt{-g}e^{-\Phi(z)}\Tr F_{MN}F_{KL}g^{MK}g^{LN}+\\
&+\frac1{k_s}\int d^5x\sqrt{-g}e^{-\Phi(z)}\bigg[ \Tr g^{MN}(D_MH)^\top(D_NH)
-M^2 \Tr HH^\top- M^2\Tr(HD^\top+H^\top D) \bigg] \nn
\end{align}
As was previously mentioned this 5D effective action draws its inspiration from
generalized sigma models including spin zero and spin one fields that
have been used in the context of strong interactions.
The normalization constants have the dimensionality $[g_5^2]=[k_s]=L^1$
to compensate that of the additional dimension. 
Following the SW holographic approach we have introduced a dilaton $\Phi(z)$.
Together with the metric $g_{MN}$ it gives the gravitational background
of a smoothly capped off spacetime.

The field strength tensor is  \be F_{MN}=(\partial_MA_N^A-\partial_NA_M^A+C^{ABC}A^B_MA^C_N)T^A,\ee
where the upper index runs through both broken and unbroken indices 
$A_M=A_M^AT^A=A_M^aT^a+A_M^i\widehat{T}^i$. This vector field is unrelated to the $W$ or
$B$ gauge bosons of the electroweak interactions.
It is dual to the vector composite 
operator  $i\Tr[T^A,s]\partial^\mu s^\top$.
$H$ is dual to the scalar composite operator $ss$. It is a matrix valued scalar field that contains the 
Goldstone bosons associated to the breaking of the global symmetry as well as other scalar fields describing perturbations in
the unbroken directions. The last term of Eqn.~(\ref{5Daction}) appears as a shifted vacuum expectation
value $H\rightarrow H+D$ and will play a crucial role in getting the phenomenological spectrum.

The dynamical breaking from $SO(5)$ to $SO(4)$ is present due to
a function $f(z)$ appearing in the nonlinear parametrization of the field $H$:
\be
H=\xi\Sigma\xi^{-1},\quad\Sigma=\begin{pmatrix}
          0_{4\times4} & 0\\
          0 & f(z)\\
         \end{pmatrix}+iT^a\sigma^a(x,z),
         \quad \xi=\exp\left(\frac{i\Pi^i(x,z)\widehat{T}^i}{\sqrt 2 f(z)}\right)
\ee
If the group elements are denoted $g \in SO(5)$ and $h\in SO(4)$, the fields transform as
\be
\xi\rightarrow\xi'= g\xi h^{-1}, \ \Sigma\rightarrow\Sigma'=h \Sigma h^{-1},\ 
H\rightarrow H'=g H g^{-1}.
\ee
The action would be fully $SO(5)$ invariant if the matrix D transformed under $SO(5)$ as $D\to g D g^{-1}$. However,
 we adopt the following form for $D$: 
\be
 D=\begin{pmatrix}
		0_{4\times4} & 0\\
		0 &  b (z)\\
		\end{pmatrix},
\ee
therefore making the last term of Eqn.~(\ref{5Daction}) only $SO(4)$ invariant. The function $b(z)$ parametrizes a soft breaking
of $SO(5)$ down to $SO(4)$ in the 5D holographic model.

Holography prescribes that every global symmetry of the 4D model comes as a gauge symmetry of its 5D dual.
Thus, to make the Lagrangian invariant under the gauge transformation
$A_M \rightarrow A'_M=gA_M g^{-1}+ig\partial_M g^{-1}$
the covariant derivative is introduced in the 5D action~(\ref{5Daction}), defined as
\be
D_MH=\partial_M H-i[A_M, H], \quad D_MH\rightarrow g D_MH g^{-1}.
\ee

We choose to work within $A_z=0$ gauge
\footnote{Doing this we depart from the studies of the holographic MCHM \cite{ACP_2005} and \cite{Falkowski2008}, where
$A_z$ is the Higgs field.},
in which the scalar kinetic term simplifies to
\be
g^{MN}(D_MH)^\top(D_NH)=g^{\mu\nu}(D_\mu H)^\top(D_\nu H)-\frac{z^2}{R^2}(\partial_z H)^\top(\partial_z H).
\ee
Expanding the scalar matrix $H$ we rewrite the  action (\ref{5Daction}) in terms of 6 scalar ($\sigma^a$), 
4 Goldstone ($\Pi^i$) and 10 vector ($A_\mu^{a/i}$) composite fields:
\begin{align}\notag
S_{5D}=&\int d^5x\sqrt{-g}e^{-\Phi(z)}\left(-\frac{1}{4g_5^2}\Tr \left(F_{\mu\nu}F_{\lambda\rho}g^{\mu\lambda}g^{\nu\rho}-
\frac{z^2}{R^2}g^{\mu\nu}\partial_z A_\mu\partial_z A_\nu\right)+\frac1{k_s}g^{\mu\nu}f^2(z)A^i_\mu A^i_\nu\right)+\\ 
\label{lagr2} &+\frac1{k_s}\int d^5x\sqrt{-g}e^{-\Phi(z)}\Bigg[
g^{\mu\nu}\partial_\mu\sigma^a\partial_\nu\sigma^a-\frac{z^2}{R^2}\partial_z\sigma^a\partial_z\sigma^a-M^2\sigma^a\sigma^a+\\
&+\frac12g^{\mu\nu}\partial_\mu\Pi^i\partial_\nu\Pi^i-\frac12\frac{z^2}{R^2}\partial_z\Pi^i\partial_z\Pi^i
-M^2f(z)b(z)\cos{\frac{\sqrt{\Pi^i\Pi^i}}{f(z)}} -\sqrt 2 f(z) g^{\mu\nu}A^i_\mu\partial_\nu \Pi^i \notag
\Bigg].
\end{align}
After imposing the $A_z=0$ gauge we still have enough gauge freedom left to set $\partial^\mu A^i_\mu =0$ and the 
mixing of the scalar $\Pi^i$ and the longitudinal part of $A^i_\mu$ can be neglected. 

The ans\"atze for the background functions $\Phi(z)$, $f(z)$ and $b(z)$ will be proposed below.

\subsection{AdS/CFT prescriptions}
Let us discuss briefly what basic holographic assumptions and prescriptions should be
taken into account.

The new strongly interacting sector being confining is supposed to have some non-Abelian gauge 
group, which could be characterized with the number of `technicolours' $N_{tc}$.
In AdS/CFT the duality is valid only in the large-$N_{tc}$ limit; however, we will relax
this condition when applying it to phenomenology.

All bulk fields are prescribed to acquire a mass following the general formula 
$M^2R^2=(\Delta-p)(\Delta+p-4)$, where $\Delta$ and $p$ are respectively a dimension 
and a spin of a dual operator \cite{Gubser1998,*Witten_1998}. Hence, the gauge fields $A^A_\mu(x,z)$ have $M^2R^2=0$ and the scalar field 
$H(x,z)$ gets $M^2R^2=-4$, the last is in fact the lowest value allowed by 
the Breitenlohner-Freedman bound.

From the 5D model given by Eqn.~(\ref{5Daction}) one can extract the n-point correlation functions 
of the composite operators. The 4D partition function in the discussed model is given analogously to any quantum field theory by
\bea
\mathcal Z_{4D}[\phi_{\mathcal O}]&=&\int [\mathcal D s]\Exp i\int d^4x [\mathcal L_{str. int.}(x)+ 
\phi_{\mathcal O \mu}^A(x)\Tr\partial^\mu s[iT^A,s](x)
+\phi_{\mathcal O}^{\alpha\beta}(x)s_{\beta\gamma} s_{\gamma\alpha}(x)] = \nn \\ 
&=&\Exp\sum\limits_q\frac1{q!}\int\prod\limits_{k=1}^qd^4x_k\langle\mathcal O_1(x_1)...\mathcal O_q(x_q)\rangle
i\phi_\mathcal O ^1(x_1)...i\phi_\mathcal O ^q(x_q),\label{Zqft}
\eea
where $\phi_{\mathcal O \mu}^A(x)$ and $\phi_{\mathcal O}^{\alpha\beta}(x)$ are the sources of the composite operators.
The AdS/CFT correspondence principle states the equivalence between the partition function in the 4D theory
and the 5D holographic effective acion when the last one is on-shell:
\be\label{cor_corresp}
\mathcal Z_{4D}[\phi_{\mathcal O}]=\Exp i S_{5D}^{on-shell}|_{\phi(x,z)\rightarrow\phi(x,z=\varepsilon)}.
\ee
Going on-shell is accompanied with setting all 5D bulk fields $\phi(x,z)$ 
to their boundary values at $z=\varepsilon$ ($\varepsilon$ being an UV regulator),
which basically coincide with the sources $\phi_\mathcal O$.
We should be more careful at this point, however. For the gauge 
fields the matching is simple $A^A_\mu(x,z)|_{z=\varepsilon}=\phi_{\mathcal O \mu}^A(x)$, while
a general scalar matrix valued field $H$ is connected to the source via
$H^{\alpha\beta}(x,z)|_{z=\varepsilon}=R^{-1}\varepsilon^{d-\Delta} \phi_{\mathcal O}^{\alpha\beta}(x)
$, where an additional length scale is introduced to have $[H]=E^1,\ [\phi_\mathcal O]=E^{d-\Delta}$.
For the degenerate situation when $\sqrt{\frac {d^2}4+M^2R^2}=0$, which is the case for $d=4,\ \Delta=2$, 
the connection contains a logarithm as well:
$H^{\alpha\beta}(x,z)|_{z=\varepsilon}=R^{-1}\varepsilon^{2} 
\ln \left(\frac\varepsilon R\right)^{2}\phi_{\mathcal O}^{\alpha\beta}(x)$ \cite{Klebanov_1999, Minces2000}.
The Green's functions can therefore be obtained by 
differentiating the 5D effective action with respect to the
sources.

Now we would like to prescribe the sources to real physical degrees of freedom, meaning the fields $\sigma$ and $\Pi$.
It is clear that quantum fluctuations of $H$ can be splitted into symmetric and antisymmetric parts
$H(x,z)|_{z=\varepsilon}=T_{sym}^iH_{sym}^i(x,\varepsilon)+T_{asym}^aH_{asym}^a(x,\varepsilon)$.
Though they do not span the ${\bf 10}_A$ and ${\bf 14}_S$ representation spaces
of $SO(5)$ (as $H$ itself does not span the whole space of $5\times 5$ matrices), they can be written in terms  
of generators forming the basis for ${\bf 10}_A$ and ${\bf 14}_S$.
Taking into account the nonlinear realization of $H$ one has an argument of the large-$N_{tc}$ limit for linearizing
the boundary value of $H$ as follows:
$H(x,z)|_{z=\varepsilon}\simeq iT^a\sigma^a(x,\varepsilon)
+ \widehat{T}_{sym}^i \Pi^i(x,\varepsilon)/\sqrt 2 ,$
where $T^a$ of Eqn.~(\ref{genso4}) form a subset in the basis of ${\bf 10}_A$ 
and $\widehat{T}_{sym}^i$ -- in the basis of ${\bf 14}_S$,
and are defined as $\widehat{T}_{sym\ \alpha\beta}^i=\frac1{\sqrt2} (\delta^i_\alpha\delta_\beta^5+\delta_\beta^i\delta_\alpha^5)$.
One may perform a similar
expansion in the sources  $\phi_{\mathcal O\ \alpha\beta}=s^a_{\mathcal O}\cdot iT^a_{\alpha\beta}
+p^{i}_{\mathcal O}\cdot \widehat{T}^i_{sym\ \alpha\beta}$.
Then it is straightforward to establish the connections between the new sources and composite operators 
related to the unbroken/broken directions:
\bea \label{scsource}
s^a_{\mathcal O}(x)=R\varepsilon^{-2}\ln^{-1} \left(\frac \varepsilon R\right)^{2} \times \sigma^a(x,\varepsilon),
\ &\text{dual to}\ \mathcal O_s^a= iT^a_{\alpha\beta}s_{\beta\gamma} s_{\gamma\alpha}(x);\\
p^{i}_{\mathcal O}(x)= R\varepsilon^{-2}\ln^{-1} \left(\frac \varepsilon R\right)^{2} \times \frac{\Pi^i(x,\varepsilon)}{\sqrt2},
\ &\text{dual to}\ \mathcal O_p^i= \widehat T^i_{sym\ \alpha\beta}s_{\beta\gamma} s_{\gamma\alpha}(x).
\eea
       
A general 5D field is a solution of a second order equation of motion (EOM) and has two modes.
The leading at small $z$ mode  gives a connection between a source at the boundary 
and a value of a field in the bulk through the bulk-to-boundary propagator. This mode should also exhibit 
enough decreasing behaviour at $z\rightarrow+\infty$ to render the on-shell action finite. 
The subleading mode provides
a set of normalizable solutions, which could be identified with a tower of physical states at the 
4D boundary.
This is the standard Kaluza-Klein (KK) mode:
the 5D field is presented as an infinite series of 4D fields weighted with the $z$-dependent profile 
functions. A boundary condition should be chosen for this profile and  
we prefer the Dirichlet one through all the paper.

It is important to understand that the 5D fields do not represent `particles'.
Therefore it should not come as a surprise the following situation 
that is one of the focal points of our work: if one sets the $SO(5)$ - breaking term $b(z)$ equal to zero, there is no
quadratic term in the 5D action Eqn. (\ref{lagr2}) for the $\Pi^i$. However, this does not mean that the particles
forming the KK tower are massless. In fact they are not. The role of the symmetry breaking
term will be to shift those masses so that the lowest lying state can be identified as a genuine Goldstone boson. 

In the following sections we show how this AdS/CFT dictionary works in practice in the
model under consideration.


\section{Unbroken generators}
\subsection{Vector fields}
For the vector fields corresponding to the unbroken generators with $a=1,...,6$, taking only transverse part, 
we get the equation of motion from the action~(\ref{lagr2})
\be\label{veom}
\left(\partial_z \frac{e^{-\Phi}}z \partial_z A^{a}_\mu-\frac{e^{-\Phi}}z \Box A_\mu^{a}\right)_\bot=0.
\ee
The dilaton background is assumed to have the standard quadratic form
$\Phi(z)=\kappa^2z^2$~\cite{SW_2006} throughout the paper. 
We perform a 4D Fourier transform  $A_\mu^a(x,z)=\int d^4q e^{iqx}A_\mu^a(q,z)$ and
focus on finding first the bulk-to-boundary propagator, which we call $V(q,z)$.
Following the holographic prescriptions, we know that
\be
A_\mu^a(q,z)=V(q,z)\phi_{\mathcal O\mu}^a(q),\quad\ V(q,\varepsilon)=1
\ee
Then, after a suitable 
change of variables $y=\kappa^2z^2$ we arrive at the following EOM:
\be
y V''(q,y)-yV'(q,y)+\frac{q^2}{4\kappa^2}V(q,y)=0
\ee
It is a particular case of the confluent hypergeometric equation (see Appendix~\ref{hyperf} 
for a review of the properties and solutions of this equation), and the solution is
\be\label{vprop}
V(q, \kappa^2z^2)=C(q)\kappa^2z^2\ _1F_1\left(-\frac{q^2}{4\kappa^2}+1,2;\kappa^2z^2\right)
+\Gamma\left(-\frac{q^2}{4\kappa^2}+1\right)\Psi\left(-\frac{q^2}{4\kappa^2},0;\kappa^2z^2\right),
\ee
where $\Psi$ is the Tricomi's function.
The second term is dominant at small $z$ and represents the propagator.

The first term gives us the tower of massive states, which could be identified with some 
physical states at the boundary. 
Normalizable solutions can only be found for  discrete values of the 4D momentum
$q^2=M_V^2(n)$ and we may identify $\left.V(q,z)\right|_{q^2=M_V^2(n)}=V_n(z)$.
Then, the KK decomposition is given as
\be A^a_\mu(q,z)=\sum\limits_{n=0}^\infty V_n(z)A_{\mu (n)}^{a}(q).\ee
The $z$ profile is determined from~(\ref{vprop}) and the spectrum can be expressed using the 
discrete parameter $n=0,1,2,...$:
\be\label{vmodes}
V_n (z)=\kappa^2z^2\sqrt\frac{2}{n+1} L_n^1(\kappa^2z^2),\quad \
M_V^2(n)=4\kappa^2(n+1),
\ee
where $L_n^m(x)$ are the generalised Laguerre polynomials. The $V_n(z)$ are subject
to the Dirichlet boundary condition and are normalized to fulfil the 
orthogonality relation:
\be\int\limits_0^\infty dz e^{-\kappa^2z^2}z^{-1}V_n(z)V_k(z)=\delta_{nk},\ee
the weight there is dictated by the form of the initial operator in Eqn.~(\ref{veom}).

\subsection{Scalar fields}
The EOM for the $\sigma(x,z)$ fields obtained from the action~(\ref{lagr2}) is:
\be\label{sc1}
\partial_z \frac{e^{-\Phi}}{z^3}\partial_z\sigma^a -\frac{e^{-\Phi}}{z^3} \Box\sigma^a
-\frac{M^2R^2}{z^5}e^{-\Phi}\sigma^a=0.
\ee
As before we take $\Phi(z)=\kappa^2z^2$. Making a 4D Fourier transformation and the substitution 
$\sigma^a(q,z)=(\kappa^2z^2)^{1-\sqrt{1+M^2R^2/4}}\widetilde{\sigma^a}(\kappa^2z^2,q)$ we arrive 
at a confluent hypergeometric equation
for $\widetilde{\sigma}^a(z,q)$, which is conveniently written in terms of $y=\kappa^2z^2$ as
\be
y\widetilde{\sigma}''_a(q,y)+\left(1-\sqrt{4+M^2R^2}-y\right)\widetilde{\sigma}'_a(q,y)
-\left(1-\frac{q^2}{4\kappa^2}-\sqrt{1+\frac{M^2R^2}4}\right)\widetilde{\sigma}_a(q,y)=0.
\ee
For the case $M^2R^2=-4$ it has the following solutions
expressed in terms of the $z$ variable
\be\label{sc2}
\sigma^a(q, z)=C_1(q)(\kappa^2z^2)\ _1F_1\left(1-\frac{q^2}{4\kappa^2},1;\kappa^2z^2\right)
+C_2(q)(\kappa^2z^2)\Psi\left(1-\frac{q^2}{4\kappa^2},1;\kappa^2z^2\right).
\ee

Now, we have two modes with the seemingly same small $z$ behaviour. This is expected in the  AdS/CFT setting 
for $\Delta=2$ scalar operators~\cite{Klebanov_1999, Minces2000}. 
However, it is known that the Tricomi's function ($\Psi$) with an integer 
second parameter exhibits a logarithmic behaviour (see Eqn.~(\ref{Triclog})).
This proves it to be the mode representing the bulk-to-boundary propagator.

To find the normalizable solutions we solve the eigenvalue problem imposing $q^2=M_\sigma^2(n)$
and demanding the Dirichlet boundary condition to be fulfilled for the eigenmodes. 
We express this solution in terms of the Laguerre polynomials as:
\be
\sigma^a(q, z)=\sum\limits_{n=0}^\infty R^{-1}\sigma_n(z)\sigma^a_{(n)}(q),\quad
\sigma_n(z)=N(\kappa z)^2L_n(\kappa^2z^2),
\ee
where $n=\frac{M_\sigma^2(n)}{4\kappa^2}-1$.
The normalization is $N^2=\frac{2}{\kappa^2}$ and 
$\int\limits_0^\infty dz e^{-\kappa^2z^2} z^{-3}\sigma_n(z)\sigma_k(z)=\delta_{nk}$. Note that
the $z$-profiles $\sigma_n(z)$ have the dimensionality $E^{-1}$, which is correct, as the dimensionality 
of the Green's function corresponding to the operator of Eqn.~(\ref{sc1}) is $E^{-4}$. Hence, we include 
the dimensionful $R$ in the KK decomposition. Compare with the discussion about the scalar sources, where
an additional $R$ also appears.

The mass spectrum is
\be
M_\sigma^2(n)=4\kappa^2(n+1+\sqrt{1+M^2R^2/4})=4\kappa^2(n+1), \ n=0,1,2...
\ee

The proposed treatment of the scalar fields is parallel to the AdS/QCD one \cite{Colangelo2008}
(for a HW alternative see \cite{DaRold_scalar})
but for the fact of having another $\Delta$ and $M^2R^2=-4$. Unlike QCD in general and
its AdS/QCD imitation, we observe a degeneracy
between the scalar and the vector fields associated to the unbroken generators. 


\section{Broken generators}
\subsection{Vector fields}
For the transverse part of the fields corresponding to the broken generators with $i=1,2,3,4$
we get the EOM:
\be
\left(\partial_z \frac{e^{-\Phi}}z\partial_z A_\mu^i-\frac{e^{-\Phi}}z \Box A_\mu^i-2
\frac{g_5^2f^2(z)}{k_s}e^{-\Phi}\frac{R^2}{z^3}A_\mu^i\right)_\bot=0.
\ee
As in the unbroken case we perform the 4D Fourier transformation and 
establish the propagation between the source and the bulk:
\be
A_\mu^i(q,z)=A(q,z)\phi_{\mathcal O\mu}^i(q),\qquad A(q,\varepsilon)=1.
\ee
Changing variables to $y=\kappa^2z^2$ again we arrive at the following EOM
\be
y A''(q,y)-yA'(q,y)+\left(\frac{q^2}{4\kappa^2}-\frac{(g_5Rf(y))^2}{2yk_s}\right)A(q,y)=0.
\ee

Now we need to choose the $z$ dependence for $f(z)$. In order to get an analytical
solution we have to assume  that either $f^2(y)\sim y$ or $f^2(y)\sim$ {\em constant}. 
The last option taken together with the boundary condition 
on $A(q,z)$ demands the implausible relation $f(y)=0$. Then, consider the linear 
ansatz\footnote{This prescription is actually natural in a sense that it implies
a quadratic scaling for $f(z)$, characteristic to chiral perturbation theory when
a non manifestly chirally invariant regulator is used.} 
$f(z)=f\cdot\kappa z$, where the constant $f$ has the dimension of mass. 
The solution of the confluent hypergeometric equation above is
\bea\label{axprop}
A(q, \kappa^2z^2)&=&C(q)\kappa^2z^2\ _1F_1\left(-\frac{q^2}{4\kappa^2}+\frac{(g_5Rf)^2}{2k_s}+1,2;\kappa^2z^2\right)+\\ 
&&+\Gamma\left(-\frac{q^2}{4\kappa^2}+\frac{(g_5Rf)^2}{2k_s}+1\right)\Psi\left(-\frac{q^2}{4\kappa^2}+\frac{(g_5Rf)^2}{2k_s},0;\kappa^2z^2\right). \nn
\eea
The second term defines the propagator, while the first for discrete values of $q^2$ and
$\left.A(q,z)\right|_{q^2=M_A^2(n)}=A_n(z)$ gives the $z$-profiles and masses of the eigenstates
\be\label{amodes}
A_n (z)=\kappa^2z^2\sqrt\frac2{n+1} L_n^1(\kappa^2z^2),\quad M_A^2(n)=4\kappa^2\left(n+1+\frac{(g_5Rf)^2}{2k_s}\right),
\quad  n=0,1,2....
\ee
We observe that the pattern of the Regge trajectory is similar to the one found for
the vector states corresponding to the unbroken generators, but the intercept is larger. That means that
these states are heavier than their unbroken counterparts. We postpone
to a latter section a tentative phenomenological discussion.

\subsection{Goldstone bosons}
The part of the action~(\ref{lagr2}) describing the $\Pi^i(x,z), \ i=1,2,3,4$ fields leads to the EOM
\be
\partial_z \frac{e^{-\Phi}}{z^3}\partial_z\Pi^i-\frac{e^{-\Phi}}{z^3} \Box\Pi^i+\frac{R^2}{z^5} \frac{b(z)}{f(z)} M^2 e^{-\Phi}\Pi^i=0.
\ee
We define $B(z)=\frac{b(z)}{f(z)}$ in substitution of
the unknown $b(z)$. Changing variables 
to $y=\kappa^2z^2$ we arrive at the following EOM
\be\label{geom}
y{\Pi_i}''(q,y)+(-1-y){\Pi_i}'(q,y)+\left(\frac{q^2}{4\kappa^2}+\frac{R^2}{4y}M^2 B(y)\right)\Pi_i(q,y)=0.
\ee
We choose a polynomial ansatz for $B (y)=\mu_1+\mu_2 y$. Addition of higher order
terms would bring us to the so-called extended confluent hypergeometric equation, solutions 
of which are not easily tractable. 

Consider the change of variables $\Pi_i(q,y)=y^\beta \widetilde{\Pi_i}(q,y)$, where 
$\beta_\pm=1\pm\sqrt{1-\frac{M^2 R^2}{4}\mu_1}=1\pm\frac12 m$ is chosen to cancel
the $y^{-1}$ terms in Eqn.~(\ref{geom}).
Then the EOM reads
\be
y\widetilde\Pi_i''(q,y)+(1\pm m-y)\widetilde\Pi_i'(q,y)
+\left(\frac{q^2}{4\kappa^2}+\frac{M^2 R^2}{4}\mu_2-(1\pm\frac12 m)\right)\widetilde\Pi_i(q,y)=0.
\ee
Due to the properties of the confluent hypergeometric functions, the  $\beta_+$ and $\beta_-$ cases are 
virtually the same; i.e.  
transforming one into another at different values of parameters (see Eqn.~(\ref{Tricrel}) in Appendix~\ref{hyperf}). 
As we expect $m$ to 
be a positive integer number, we choose the solution:
\begin{align}\notag\Pi^i(q,\kappa^2z^2)&=C_1(q)(\kappa^2z^2)^{1+\frac12 m}
\ _1F_1\left(1+\frac12 m-\frac{q^2}{4\kappa^2}-\frac{M^2 R^2}{4}\mu_2,1+m;\kappa^2z^2\right)+ \nn \\ 
&+C_2(q)(\kappa^2z^2)^{1+\frac12 m}
\Psi\left(1+\frac12 m-\frac{q^2}{4\kappa^2}-\frac{M^2 R^2}{4}\mu_2,1+m;\kappa^2z^2\right).\label{gprop}
\end{align}

Looking for a normalizable solution for discrete values
$q^2=M_\Pi^2(n)$ we arrive at the KK decomposition with the specific 5D profiles
\be
\Pi^i(q,z)=\sum\limits_{n=0}^\infty R^{-1}\Pi_n(z)\Pi^i_{(n)}(q),\quad
\Pi_n(z)=N(\kappa z)^{2+m}L_n^m(\kappa^2z^2),
\ee
where $n=-1-\frac12 m+\frac{M_\Pi^2(n)}{4\kappa^2}+\frac{M^2 R^2}{4}\mu_2,\ m=\sqrt{4-M^2 R^2\mu_1}$.
The normalization is analogous to that of $\sigma(x,z)$ fields: $N^2=\frac{2}{\kappa^2}$ and 
$\int\limits_0^\infty dz e^{-\kappa^2z^2}z^{-3}\Pi_n(z)\Pi_k(z)=\delta_{nk}$.

Thus, we obtain the following mass spectrum depending on the parameters $m$ (or $\mu_1$) and 
$\mu_2$ of the ansatz:
\be
M_\Pi^2(n)=4\kappa^2\left(n+1+\frac12 m-\frac{M^2 R^2}{4}\mu_2\right).
\ee

Having found the general solution, we focus on the case of $M^2R^2=-4$ and fix the values of
$\mu_1$ and $\mu_2$.
To determine $\mu_1$ let us turn to the bulk-to-boundary propagator connecting the 5D bulk value
to the source; its boundary
behaviour is fixed by the AdS/CFT prescription:
\be
\Pi^i(q,\kappa^2z^2)=\Pi^{prop}(q^2/\kappa^2,\kappa^2z^2)p^i_\mathcal O (q) \ \xrightarrow{z\rightarrow\varepsilon}\ 
\Pi^i(q,\kappa^2\varepsilon^2)=\sqrt 2\varepsilon^2R^{-1} \ln\left(\frac{\varepsilon} R\right)^2 
p^i_\mathcal O (q).
\ee 
Taking into account that
$\Pi^{prop}(q^2/\kappa^2,\kappa^2z^2)$ must be well-behaved at $z\rightarrow+\infty$, we conclude that only 
the $\Psi$-function term in Eqn.~(\ref{gprop}) with $m=0$ (equivalent to $\mu_1=-1$) can provide the correct solution.
As now we have the value of $m$ fixed, the eigenmodes
have the $z$-profiles and the spectrum
\be
\Pi_n(z)=\sqrt{\frac{2}{\kappa^2}}(\kappa z)^{2}L_n(\kappa^2z^2), 
\quad M_\Pi^2(n)=4\kappa^2\left(n+1+\mu_2\right),\quad n=0,1,2...
\ee
By choosing $\mu_2=-1$ we obtain the result that for $n=0$,
$M_\Pi^2(0)=0$. This is remarkable as it provides us with a multiplet of massless Goldstone bosons, a 
result that is not easy to get in the holographic approach or in the Regge theory. If we look at the origin of this more
closely we see that it is due to the combination of the two terms contributing to the mass in Eqn. (\ref{5Daction}). Note that the term linear
in $H$ only contributes to the symmetric part of this matrix valued field ($H$ is not in an irrep of $SO(5)$).


\section{Correlation functions}
\label{sec-corr}
In this section we summarize the two-point correlation functions derived in the four channels just discussed.
\subsection{Unbroken generators}
Following the holographic prescriptions given by Eqns. (\ref{Zqft}) and (\ref{cor_corresp}) 
we define the vector correlation function as
\be
\langle \mathcal O_\mu^a(q) \mathcal O_\nu^b(p)\rangle=\delta(p+q)\int d^4x e^{iqx}\langle \mathcal O_\mu^a(x) \mathcal O_\nu^b(0)\rangle
=(-i)^2\frac\delta{\delta A^a_{\mathcal O\mu}(q)}\frac\delta{\delta A^b_{\mathcal O\nu}(p)}iS_{5D}^{on-shell}.
\ee
If $\mathcal O_\mu^a(x)$ is proportional to a conserved current (which is the case), the correlator
should be transverse and one can consider only the meaningful part $\Pi(q^2)$ without Lorentz indices:
\be
i\int d^4x e^{iqx}\langle \mathcal O_\mu^a(x) \mathcal O_\nu^b(0)\rangle=\delta^{ab}\left(\frac{q_\mu q_\nu}{q^2}-\eta_{\mu\nu}\right)\Pi_{unbr}(q^2).
\ee
We note that $\Pi_{unbr}(q^2)$ is generically subject to short distance ambiguities. They are encoded in
constants $C_0$ and $C_1$ in the polynomial $C_0+C_1 q^2$ (see e.g. \cite{REINDERS1985, Afonin2006}).
 
Going on-shell in the Lagrangian~(\ref{lagr2}) we find $\Pi_{unbr}(q^2)$ to be 
\be\label{unbrA}
\Pi_{unbr}(q^2)= \frac{R}{g_5^2}\left.\left[\frac{e^{-\Phi(z)}V(q,z)\partial_z V(q,z)}{z}\right]\right|_{z=\varepsilon},
\ee
and substituting the propagator from Eqn.~(\ref{vprop}) we get
\be\label{unbrA1}
\Pi_{unbr}(q^2)=-\frac{R}{2g_5^2} q^2 \left(\ln \kappa^2\varepsilon^2+2\gamma_E+\psi\left(-\frac{q^2}{4\kappa^2}+1\right)\right),
\ee
where $\gamma_E$ is the Euler-Mascheroni constant and $\psi(a)$ is the digamma function.
In the limit of $Q^2=-q^2\rightarrow\infty$ we have (using the Stirling's approximation for $\psi$)
\be
\Pi_{unbr}(Q^2)=\frac{R}{2g_5^2} Q^2 \left(\ln(Q^2\varepsilon^2)-\ln 4+2\gamma_E+\frac{2\kappa^2}{Q^2}-
\frac{4\kappa^4}{3Q^4}+\mathcal O\left(\frac1{Q^6}\right)\right).
\ee
On the other hand, one can evaluate the leading order Feynman diagrams at the microscopic level,
that is done in Appendix~\ref{loops}.
Matching  the coefficient of the logarithm with the  perturbative loop expression 
we get the $\frac{g_5^2}{R}$ coefficient fixed (see Eqn.~(\ref{g5fix})). Note the presence of a $1/Q^2$ term
that is absent in QCD in the chiral limit \cite{Afonin2006}.

To have a correlator in a phenomenological form of the resonance decomposition we should perform a 
decomposition of the digamma function
in (\ref{unbrA1}) leading to
\be\label{unbrA2}
\Pi_{unbr}(q^2)=-\frac{R}{2g_5^2}  \left(\ln \kappa^2\varepsilon^2+\gamma_E\right)q^2
-\frac{2\kappa^2R}{g_5^2}  \sum\limits_{n=0}^\infty\frac{q^4}{M^2_V(n)(q^2-M^2_V(n))}.\ee

The first bracket would correspond to the short distance ambiguity mentioned above (constant $C_1$). 
The second term is a well convergent sum over the resonances.

Alternatively, we could have calculated the same two-point function introducing the resonances at an earlier stage.
As we have found a tower of normalized massive eigenstates~(\ref{vmodes}) it is straightforward to construct 
the Green's function of the 
Sturm-Liouville operator appearing in Eqn.~(\ref{veom}):
$G(q,z,z')
=\sum\limits_{n=0}^\infty\frac{V_n(z)V_n(z')}{q^2-M^2_V(n)}$. 
Inserting the relation connecting the value of the 5D field to the Green's function
$V(q,z)=-\lim\limits_{z'\rightarrow0}\frac{e^{-\Phi(z')}}{z'}V(q,z')\partial_{z'} G(q,z,z')$ 
into Eqn.~(\ref{unbrA}) we find
\be
\Pi_{unbr}(q^2)=-\frac{R}{g_5^2}\lim\limits_{z,z'\rightarrow0}\frac1{zz'}\partial_{z}\partial_{z'}G(q,z,z')
=-\frac{R}{g_5^2}\sum\limits_{n=0}^\infty\frac{(V'_n(\varepsilon)/\varepsilon)^2}{q^2-M^2_V(n)}.
\ee
After the substitution of the normalized modes from Eqn.~(\ref{vmodes}) this gives the correlator
\begin{align}
\Pi_{unbr}(q^2)&=-\frac{R}{g_5^2}\sum\limits_{n=0}^\infty\frac{8\kappa^4(n+1)}{q^2-M^2_V(n)}=\\ \label{unbrA3}
&=-\frac{2\kappa^2R}{g_5^2} \sum\limits_n\frac{q^4}{M^2_V(n)(q^2-M^2_V(n))}+q^2\sum\limits_n\frac{2\kappa^2R/g_5^2}{M^2_V(n)}+\sum\limits_n\frac{2\kappa^2R}{g_5^2}.
\end{align}
The first sum in (\ref{unbrA3}) is the one left after making the proper subtractions,
and therefore the one informative for the resonance description of the two-point
function, and coincides with the sum in Eqn.~(\ref{unbrA2}). The convergent correlator is 
\be\label{FV}
\widehat\Pi_{unbr}(Q^2)=\sum\limits_{n=0}^\infty\frac{Q^4F_V^2}{M^2_V(n)(Q^2+M^2_V(n))},\quad F_V^2=\frac{2R\kappa^2}{g_5^2}.
\ee
The second sum (the term proportional to $q^2$) in (\ref{unbrA3}) corresponds to the subtraction constant $C_1$, which was 
already determined in Eqn.~(\ref{unbrA2}). Thus, it seems necessary to match the two expressions of $C_1$. 
This imposes a connection between the maximum number of resonances $N_{max}$ and the UV regulator $\varepsilon$:
 $\ln N_{max}=-2\gamma_E-\ln \kappa^2\varepsilon^2$. This relation should be interpreted as being only 
 really meaningful at the leading order (i.e. the constant non-logarithmic part cannot be
 determined by this type of heuristic arguments).
The last sum in (\ref{unbrA3}) actually corresponds to a quadratic (and potentially a subleading logarithmic) divergence
as it behaves as $N_{max}^2$ if we sum up a finite number of resonances. Therefore, it can be eliminated 
by redefining the subtraction constant $C_0$ previously discussed.

It should not come as a surprise that the subtractions required are different when deriving the correlator
in two different ways as this is a divergent quantity, ill-defined at short distances and reordering of the
manipulations may lead to different results. It is fundamental, however, that the ambiguities should be limited to
the form $C_0 +C_1 q^2$.

We may also consider the two-point correlation function of the scalar operators, it is determined as the second 
variation with respect to the proper sources of the 5D action on the boundary:
\be
\langle \mathcal O_s^a(q) \mathcal O_s^b(p)\rangle=\delta(p+q)\int d^4x e^{iqx}\langle \mathcal O_s^a(x) \mathcal O_s^b(0)\rangle
=(-i)^2\frac\delta{\delta  s^a_{\mathcal O}(q)}\frac\delta{\delta  s^b_{\mathcal O}(p)}iS_{5D}^{on-shell},
\ee
and the quantity $\Pi_S(q^2)$ is introduced as
\be
i\int d^4x e^{iqx}\langle \mathcal O_s^a(x) \mathcal O_s^b(0)\rangle=\delta^{ab}\Pi_S(q^2).
\ee

The scalar case is quite different from the vector one, mostly because of the dimensionality of the scalar operator
under investigation.
For the sake of clarity we propose to keep all 4D integrals
and seek for $\Pi_S(q^2)$ as a part of the on-shell boundary value of the $\sigma$ action
\be
I_{\partial AdS}=\frac12\int d^4xd^4y s^a_\mathcal O(x) s^b_\mathcal O(y)\langle \mathcal O_s^a(x)\mathcal O_s^b(y)\rangle.
\ee
We start with $I_{\partial AdS}=-\left.\frac1{k_s}\int d^4x \frac{R^3}{\varepsilon^3}\sigma(x,z)\partial_z\sigma(x,z)\right|_{z=\varepsilon}$,
where $\sigma(x,z)$ is the inversed Fourier transformation
of the regular at $z\rightarrow\infty$ branch of the solution of EOM ($\Psi$ function in Eqn.~(\ref{sc2})). After substitution
and proper $\varepsilon$ expansion we get
\be
I_{\partial AdS}=-\frac{2R}{k_s}\int d^4xd^4ys^a_\mathcal O(x)s^b_\mathcal O(y)\int\frac{d^4q}{(2\pi)^4}
e^{-iq(x-y)}\left[\ln\frac{\varepsilon^2}{\kappa^2R^4}-\psi\left(1-\frac{q^2}{4\kappa^2}\right)
-2\gamma_E\right]\delta^{ab},
\ee
where the scalar sources are defined following Eqn.~(\ref{scsource}).
Then we have
\be\label{sccor}
\Pi_S(q^2)=-\frac{4R}{k_s}\left[\ln\frac{\varepsilon^2}{\kappa^2R^4}-2\gamma_E-\psi\left(1-\frac{q^2}{4\kappa^2}\right)
\right].
\ee
In the $Q^2\rightarrow\infty$ limit the following expansion is valid:
\be
\Pi_S(q^2)=-\frac{4R}{k_s}\left[-\ln\frac {Q^2R^4}{4\varepsilon^2}-2\gamma_E-\frac{2\kappa^2}{Q^2}+\frac43\frac{\kappa^4}{Q^4}+\mathcal O\left(\frac1{Q^6}\right)\right].
\ee
Once again, we may fix the $\frac{R}{k_s}$ coefficient at the large $Q^2$ logarithm calculating 
the loops with fundamental fields, see Eqn.~(\ref{ksfix}).

\subsection{Broken generators}
The two-point correlation functions of the vector operators corresponding to the broken 
generators are defined the same way
as the unbroken ones with a change $a,b\rightarrow i,j$.

For the vector correlator
$i\int d^4x e^{iqx}\langle \mathcal O_\mu^i(x) \mathcal O_\nu^j(0)\rangle=\delta^{ij}(\frac{q_\mu q_\nu}{q^2}-\eta_{\mu\nu})\Pi_{br}(q^2)
$ we find an expression:
\be
\Pi_{br}(q^2)= \frac{R}{g_5^2}\left.\left[\frac{e^{-\Phi(z)}A(q,z)\partial_z A(q,z)}{z}\right]\right|_{z=\varepsilon}.
\ee
Taking the propagator from Eqn.~(\ref{axprop}) we get
\be \label{brA}
\Pi_{br}(q^2)=-\frac{R}{2g_5^2} q^2 \left(1-\frac{2(g_5Rf\kappa)^2}{q^2k_s}\right)\left(\ln \kappa^2\varepsilon^2+2\gamma_E+\psi\left(-\frac{q^2}{4\kappa^2}+1
+\frac{(g_5Rf)^2}{2k_s}\right)\right).
\ee
We expect to find a "pion" pole in the $q^2\rightarrow0$ expansion, which is indeed there
\begin{align}\label{brA1}
\lim\limits_{q^2\rightarrow0}&\Pi_{br}(q^2)=-\frac{R}{2g_5^2} q^2 \bigg[\ln\kappa^2\varepsilon^2+2\gamma_E+\psi\left(1+\frac{(g_5Rf)^2}{2k_s}\right)+
\frac{(g_5Rf)^2}{2k_s}\psi_1\left(1+\frac{(g_5Rf)^2}{2k_s}\right)-\notag
\\&-\frac{4\kappa^2}{q^2}
 \frac{(g_5Rf)^2}{2k_s}\left(\ln\kappa^2\varepsilon^2+2\gamma_E+\psi\left(1+\frac{(g_5Rf)^2}{2k_s}\right)\right)+\mathcal O(q^4)\bigg]
 =-F^2+\widetilde\Pi_{br}(q^2),
\end{align}
where have defined the `pion decay constant' as 
\be
\label{brA2}
F^2=-\frac{\kappa^2f^2R^3}{k_s}\left(\ln\kappa^2\varepsilon^2+2\gamma_E+\psi\left(1+\frac{(g_5Rf)^2}{2k_s}\right)\right).\ee
We presume this $F$ to be the quantity defining the scale of the 
global symmetry breaking in the 4D theory.

In the limit of $Q^2\rightarrow\infty$ we have an expression which expectedly coincides with the unbroken case if
$f=0$
\begin{align}\label{brA3}
\notag\Pi_{br}(Q^2)=\frac{R}{2g_5^2} Q^2 &\Bigg(\ln(Q^2\varepsilon^2)-\ln 4+2\gamma_E+\frac{2\kappa^2}{Q^2}
\left[1+\frac{(g_5Rf)^2}{k_s}(1+\ln(Q^2\varepsilon^2)-\ln4+2\gamma_E)\right]\\ 
&+\frac{\kappa^4}{Q^4}\left[-\frac{4}3+2 \frac{(g_5Rf)^4}{k_s^2}\right]+\mathcal O\left(\frac1{Q^6}\right)\Bigg).
\end{align}

Using the series form of the digamma function in Eqn.~(\ref{brA}) we get:
\be \label{brA4}
\Pi_{br}(q^2)=-\sum\limits_n\frac{q^4 F_A^2(n)}{M^2_A(n)(q^2+M^2_A(n))}-F^2
 -q^2\left(\frac{R}{2g_5^2}  \left(\ln \kappa^2\varepsilon^2+\gamma_E\right)+
\sum\limits_n\left(\frac{F^2_V}{M^2_V(n)}-\frac{F^2_A(n)}{M^2_A(n)}\right)\right),
\ee
where we have introduced 
\be F_A^2(n)=\frac{2R\kappa^2}{g_5^2}\frac{n+1}{n+1+\frac{(g_5Rf)^2}{2k_s}}.\ee

As in the unbroken case we can construct an alternative expression
using the eigenstates of Eqn.~(\ref{amodes}), which coincide with those of Eqn.~(\ref{vmodes})
but have different masses:
\begin{align} 
&\Pi_{br}(q^2)=-\frac{R}{g_5^2}\sum\limits_n\frac{8\kappa^4(n+1)}{q^2-M^2_A(n)}=\\ \label{brA5}
&=-\sum\limits_n\frac{q^4 F^2_A(n)}{M^2_A(n)(q^2-M^2_A(n))}
-F^2
+\sum\limits_n \frac{2\kappa^2R}{g_5^2}+q^2\sum\limits_n\frac{F^2_A(n)}{M^2_A(n)}.
\end{align}

Now the first two sums of (\ref{brA5}) are the ones meaningful for the resonance description of the two-point function
\begin{equation}\label{brA6}
\widehat\Pi_{br}(Q^2)=\sum\limits_n\frac{Q^4 F_A^2(n)}{M^2_A(n)(Q^2+M^2_A(n))}-F^2,\quad 
F^2=\frac{2R\kappa^2}{g_5^2}\sum\limits_n\frac{\frac{(g_5Rf)^2}{2k_s}}{n+1+\frac{(g_5Rf)^2}{2k_s}},
\end{equation}
and the last two correspond to quadratic and logarithmic subtractions, respectively.

The requirement that the two expressions for $F^2$ ((\ref{brA2}) and (\ref{brA6})) and the different $q^2$ 
terms of (\ref{brA4}) and (\ref{brA5}) coincide
demands the fulfilment of the relation
$\ln N_{max}=-2\gamma_E-\ln \kappa^2\varepsilon^2$ found previously in the unbroken case. The $q^2$ 
subtraction shows that the
renormalization ambiguity involved in the constant $C_1$ is of ultraviolet origin as this
is independent whether symmetries are broken or unbroken. The determination of $F^2$ in (\ref{brA5})
is straightforward as soon as we expect that the `quadratic' term $\sum\limits_n \frac{2\kappa^2R}{g_5^2}$
to be subtracted (i.e. associated to $C_0$) is the same both in the broken and unbroken channels. Again
this is a reflection of the UV nature of the ambiguity.

Similar to the unbroken scalar case we also get the two-point correlation function of the scalar operators
corresponding to the broken directions and dual to the 5D Goldstone fields:
\begin{align}
&i\int d^4x e^{iqx}\langle \mathcal O_p^i(x) \mathcal O_p^j(0)\rangle=\delta^{ij}\Pi_G(q^2),\\\label{goldcor}
&\Pi_{G}(q^2)=-\frac{4R}{k_s}\left[\ln\frac{\varepsilon^2}{\kappa^2R^4}-2\gamma_E-\psi\left(-\frac{q^2}{4\kappa^2}\right)
\right],
\end{align}
where the poles of the digamma function prove that we have a massless state in a tower of resonances.
In fact, the correlator (\ref{goldcor}) and the scalar correlator of Eqn.~(\ref{sccor}) coincide but
for the massless pole: $\Pi_S(q^2)-\Pi_{G}(q^2)=-\frac{4R}{k_s}\frac{4\kappa^2}{q^2}$. This behaviour
does not match the QCD expectation of vanishing as $1/Q^4$ when $Q^2\rightarrow\infty$.
         
\section{Inclusion of the SM gauge bosons}\label{SMincl}
In this section  we develop a possible scenario of the SM group gauging via introducing the SM  gauge 
bosons as non-dynamical fields in a bulk of AdS coupled to to the composite fields in a particular way.
That does not affect our previous computations of the spectra of composite states and presents a self-consistent
way to involve the weak scale physics into the holographic model.

Taking into account that the SM gauge bosons are weakly coupled fields, they do 
not seem to admit a holographic treatment and the only guiding principle at hand is to use gauge invariance (with respect to the electroweak group). 
That does uniquely fix the way one can extend the 5D covariant derivative of the Lagrangian (\ref{5Daction}) with the 
SM gauge fields. It is convenient to include them in an redundant $SO(4)'$ multiplet ${X}_\mu(x)$:
\be\label{covD}
D_\mu H(x,z)=\partial_\mu H(x,z)-i[A_\mu(x,z), H(x,z)]-i[\widetilde{X}_\mu(x), H(x,z)],
\ee
where the tilde means that the $X_\mu^a\ (a=1,...,6)$ fields come with the rotated generators 
$\widetilde{X}^a_\mu=X_\mu^{a}T^{a}(-\theta)=X_\mu^{a}r^{-1}(\theta)T^{a}(0)r(\theta)$. 
No dependence on the $z$ direction is assumed for $X_\mu$. 
Note that in spite of $A_\mu$ and $\widetilde{X}_\mu$ having similar couplings to the matter field $H$
it is not possible to eliminate $X_\mu$ by a field redefinition as it does not appear elsewhere in the action.

The above modification results in the following additional quadratic terms to the 5D action
(we enumerate the six generators in a way that the first three components of $X_\mu^a$
may be named $X_\mu^{L\ \alpha}$ and the last three $X_\mu^{R\ \alpha}$):
\begin{align}\label{EWquad}
\varDelta S_{5D}=\frac1{k_s}\int& d^5x\sqrt{-g}e^{-\Phi(z)}g^{\mu\nu}\sum_{\alpha=1}^3\bigg(
\frac12 f^2(z)\sin^2\theta (X^{L\ \alpha}_\mu-X^{R\ \alpha}_\mu) (X^{L\ \alpha}_\nu-X^{R\ \alpha}_\nu)-\notag\\
&-f(z)\sin\theta\partial_\mu\Pi^\alpha(X^{L\ \alpha}_\nu-X^{R\ \alpha}_\nu)+ \sqrt2 f^2(z)\sin\theta  A^{br\ \alpha}_\mu(X^{L\ \alpha}_\nu-X^{R\ \alpha}_\nu)
\bigg).
\end{align}
Other terms, with three or more fields, result from the implementation of the extended derivative (\ref{covD}) as well.
However, these will not be needed for the present discussion. 

Eventually, a  kinetic term needs to be added to obtain the propagation of physical gauge bosons
but it still would not appear in a combination that would allow $X_\mu$ elimination via a field redefinition (a 
situation similar to what happens in phenomenological Lagrangians involving spin one resonances and gauge fields
\cite{ECKER1989, *DAMBROSIO2006}). Gauge invariance allows for terms such as $\Tr\,X_{MN}X^{MN}$, where $X_{MN}$ is the
electroweak field strength providing the gauge boson kinetic term. There are other possibilities too, such as
$\Tr\, F_{MN}\xi X^{MN}\xi^{-1}$. The holographic principle does not provide information on the coefficient of this 
operator.

Intending to gauge $SU(2)\times U(1)$ we should eliminate the excessive degrees of freedom in ${\mathcal H}^\prime$. 
We choose the left subgroup as the one where the SM gauge fields get a mass; then we should set
$X_\mu^{R\ 1}=X_\mu^{R\ 2}=0$ and rename $X_\mu^{L\ \alpha}=\frac g{\sqrt2} W_\mu^{\alpha}$
and $X_\mu^{R\ 3} =\frac {g'}{\sqrt2} B_\mu$, implying that $X_\mu^{L\ 3}-X_\mu^{R\ 3}=\frac{\sqrt{g^2+g'^2}}{\sqrt2}Z_\mu$
and the massless photon corresponds to the orthogonal combination $X_\mu^{L\ 3}+X_\mu^{R\ 3}=\frac{\sqrt{g^2+g'^2}}{\sqrt2}\gamma_\mu$.



We focus on getting the mass of the gauge fields, thus there is no necessity to consider the
mixing terms of Eqn.~(\ref{EWquad}) (essentially as they would give one-point reducible contribution to the self-energies).
The quadratic boundary action has the following mass terms after integrating over the bulk coordinate $z$ in Eqn.~(\ref{EWquad}):
\begin{align}
\varDelta S|_{\partial AdS}=\int d^4q\ \eta^{\mu\nu}\left[ \frac12
\Xi g^2(W_\mu^{1}W_\nu^{1}+W_\mu^{2}W_\nu^{2})+ \frac12\Xi (g^2+g'^2)Z_\mu Z_\nu\right],
\end{align}
with
\be\label{Wmass}
\Xi=\frac{(fR\kappa)^2}4 \frac R{k_s}\sin^2\theta\Gamma\left(0,\frac{\kappa^2}{\Lambda_\text{cut-off}^2}\right).
\ee
In Eqn.~(\ref{Wmass}) $\Gamma\left(0,\frac{\kappa^2}{\Lambda_\text{cut-off}^2}\right)$ 
is the upper incomplete gamma function, the lower limit is $z=\varepsilon=1/\Lambda_\text{cut-off}$.
The three eaten Goldstone bosons give masses to the $W_\mu^{1},\ W_\mu^{2}$ and $Z_\mu$. 
The fourth composite Goldstone boson becomes the Higgs boson, following the ideas behind 
composite Higgs models. The role of misalignment 
is obvious as $\theta=0$ cancels the effect and leads to the decoupling of the composite sector 
from the low-energy (SM) physics.

On the other hand, in the effective Lagrangian~(\ref{lagr1}) a particular $SU(2)'\times U(1)'\subset SO(4)'$ is already gauged
because only the SM fields $W_\mu^{a_L}$ and $B_\mu$  couple to the
 currents of the strongly interacting sector. 
 These are the same vectorial currents that are holographically connected to the vector composite fields. 
Hence, we may include to the 4D partition function $\mathcal Z_{4D}[\phi_\mathcal O]$ the following terms
quadratic in natural sources $W$ and $B$:
${W}^\mu\langle \widetilde J^L_\mu(q) \widetilde J^L_\nu(-q)\rangle {W}^\nu $,
${W}^\mu  \langle \widetilde J^L_\mu(q) \widetilde J^R_\nu(-q)\rangle {B}^\nu $,
${B}^\mu  \langle \widetilde J^R_\mu(q) \widetilde J^R_\nu(-q)\rangle {B}^\nu $.
Precisely, the relevant correlators are defined as follows:
\begin{align}
i\int d^4xe^{iqx} \langle \widetilde J^{a_L}_\mu(x) \widetilde J^{b_L}_\nu(0)\rangle&=\delta^{a_Lb_L}\frac{g^2}2
\left(\frac{q_\mu q_\nu}{q^2}-\eta_{\mu\nu}\right)\Pi_{LL}(q^2),\\
i\int d^4xe^{iqx} \langle \widetilde J^{a_R}_\mu(x) \widetilde J^{b_R}_\nu(0)\rangle&=\delta^{a_Rb_R}\frac{g'^2}2 
\left(\frac{q_\mu q_\nu}{q^2}-\eta_{\mu\nu}\right)\Pi_{RR}(q^2),\\ \label{LRdef}
2i\int d^4xe^{iqx} \langle \widetilde J^{a_L}_\mu(x) \widetilde J^{b_R}_\nu(0)\rangle&=
\delta^{a_Lb_R}\frac{gg'}2\left(\frac{q_\mu q_\nu}{q^2}-\eta_{\mu\nu}\right)\Pi_{LR}(q^2).
 \end{align}
 
 As the relation between the rotated operators and the unrotated ones is known $\widetilde J^{a}=r(\theta)J^{a}r^{-1}(\theta)$,
it is straightforward to express the aforementioned two-point functions in terms of the correlators
of Section~\ref{sec-corr}. $\Pi_{LL}$ and $\Pi_{RR}$ are expectedly equal:
 \be
 \Pi_{diag}(q^2)=\Pi_{LL}(q^2)=\Pi_{RR}(q^2)=\frac{1+\cos^2\theta}2 \Pi_{unbr}(q^2)+\frac{\sin^2\theta}2 \Pi_{br}(q^2),
 \ee
while the $\Pi_{LR}(q^2)$ has a different value and will be analysed later on.

The relevant quadratic contribution of the gauge bosons to the 4D partition function is:
\begin{align}
\notag
\varDelta \ln \mathcal Z_{4D}=\int d^4q\bigg[&\left(\frac{q^\mu q^\nu}{q^2}-
\eta^{\mu\nu}\right)\frac{1}{4}\Pi_{diag}(q^2)(g^2{W}_\mu^{1}{W}_\nu^{1}+g^2{W}_\mu^2{W}_\nu^2+g^2{W}_\mu^3{W}_\nu^3+g'^2{B}_\mu{B}_\nu)+
\\ \label{gaugedL}
+&\left(\frac{q^\mu q^\nu}{q^2}-\eta^{\mu\nu}\right)\frac{1}4 \Pi_{LR}(q^2) gg'{W}_\mu^{3}{B}_\nu\bigg],
\end{align}
or for the gauge bosons in the physical basis:
\begin{align}
\notag
\varDelta \ln \mathcal Z_{4D}&=\int d^4q \left(\frac{q^\mu q^\nu}{q^2}-
\eta^{\mu\nu}\right) \bigg[\frac{1}{4}\Pi_{diag}(q^2)g^2({W}_\mu^{1}{W}_\nu^{1}+{W}_\mu^2{W}_\nu^2)
+\frac{g^2+g'^2}{8}\Pi_{unbr}(q^2)\gamma_\mu\gamma_\nu+\\
&+\frac{g^2+g'^2}{8}\left(\cos^2\theta\Pi_{unbr}(q^2)+\sin^2\theta\Pi_{br}(q^2)\right)Z_\mu Z_\nu
\bigg]. 
\end{align}
This expression plays a role of an effective action containing the self-energies of the electroweak bosons.
One can follow the corresponding SM masses (considering $v=F\sin\theta$) from a part which is constant 
in $q^2 \rightarrow 0$ limit:
\be
M^2_W=\frac{g^2}4\sin^2\theta F^2,\ M^2_Z=\frac{g^2+g'^2}4\sin^2\theta F^2,\ M^2_\gamma=0. \label{Gmasses}
\ee

Thus we come to have two relations for the $W$ mass, one from Eqn.~(\ref{Wmass}) and another from Eqn.~(\ref{Gmasses}):
\begin{align}\label{MW1}
&M^2_W=- \frac{(gfR\kappa)^2}4 \frac R{k_s}\sin^2\theta\left(\ln\frac{\kappa^2}{\Lambda_\text{cut-off}^2}+2\gamma_E+\psi\left(1+\frac{(g_5Rf)^2}{2k_s}\right)\right),
\\ \label{MW2}
&M^2_W=\frac{(gfR\kappa)^2}4 \frac R{k_s}\sin^2\theta\Gamma\left(0,\frac{\kappa^2}{\Lambda_\text{cut-off}^2}\right).
\end{align}
They are not the same, and in fact agree only with logarithmic accuracy as
$\Gamma\left(0,\frac{\kappa^2}{\Lambda_\text{cut-off}^2}\right)$ is dominated by $-\ln\frac{\kappa^2}{\Lambda_\text{cut-off}^2}$.
Without doubt we regard the first relation~(\ref{MW1}), 
derived from the expression for $F^2$ that uses current-algebra reasoning, as being more accurate, consistent with the SM and holographically substantiated.
As well because Eqn.~(\ref{MW2}) depends on
our choice of the  $z$-profile for $X^a_\mu(x,z)$ in (\ref{covD}). On
the other hand, it is encouraging that the two expressions are quite similar. In either case, the mixing with vector
resonances has been neglected.

We will return to discuss what constraint the  $W$ mass~(\ref{MW1}) poses on
the model parameters later, in Section~\ref{phen}.


\section{Left-Right correlator and sum rules}
 The oblique corrections to the SM physics~\cite{AltarelliB, PT} are defined
 to follow the new physics contributions to the
 vacuum polarization amplitudes (new massive resonances in the loops). The $S$ and $T$ parameters of Peskin
 and Takeuchi~\cite{PT} are the most relevant for the discussion of the composite Higgs models.
 However, due to the custodial symmetry of the strongly interacting sector the tree-level correction to the $T$ parameter vanishes in the
$SO(5)\rightarrow SO(4)$ model under consideration. The analysis of the NLO corrections does not seem to be well-motivated
as they are suppressed in the large-$N_{tc}$ limit, where the holographic description is valid. 
Thus, we focus on the $S$ parameter connected to the $\Pi_{LR}(q^2)$.
 
The left-right two-point function has its meaning only when the SM gauge fields are introduced.
Following Eqns. (\ref{LRdef}) and (\ref{gaugedL}) we define it as
\be
2i\int d^4xe^{iqx} \langle \widetilde J^{3_L}_\mu(x) \widetilde J^{3_R}_\nu(0)\rangle=
2\frac\delta{\delta W_{3}}\frac\delta{\delta B}\varDelta \ln \mathcal Z_{4D}
=\frac{gg'}2\left(\frac{q_\mu q_\nu}{q^2}-\eta_{\mu\nu}\right)\Pi_{LR}(q^2).
\ee
For further analysis we express it in terms of the correlators calculated in Section~\ref{sec-corr}:
 \begin{align}\notag
\Pi_{LR}(q^2)=&\sin^2\theta \left(\Pi_{unbr}(q^2)-\Pi_{br}(q^2)\right)=\\\label{LR0}
=&-\frac{R}{2g_5^2}q^2\sin^2\theta\bigg[\psi\left(1-\frac{q^2}{4\kappa^2}\right)-\psi\left(1-\frac{q^2}{4\kappa^2}
+\frac{(g_5Rf)^2}{2k_s}\right)+\\&+\frac{4\kappa^2}{q^2} \frac{(g_5Rf)^2}{2k_s}\left(\ln\kappa^2\varepsilon^2+
2\gamma_E+\psi\left(1-\frac{q^2}{4\kappa^2}+\frac{(g_5Rf)^2}{2k_s}\right)\right)\bigg].\notag
\end{align}
 
 First, we define $L_{10}$, the coefficient of the chiral effective electroweak Lagrangian
 $L_{10}=\left.\frac{d}{dQ^2}\frac{\Pi_{LR}(Q^2)}4\right|_{Q^2=0}$, that turns out to be equal to
\be
 L_{10}=-\frac{R\sin^2\theta}{8g_5^2}\left[\gamma_E+\psi\left(1+\frac{(g_5Rf)^2}{2k_s}\right)
 +\frac{(g_5Rf)^2}{2k_s}\psi_1\left(1+\frac{(g_5Rf)^2}{2k_s}\right)\right].
\ee
Up to a constant it coincides with the $S$ parameter of Peskin-Takeuchi:
$S= -16\pi L_{10}$, so
 \be \label{Spar1}
 S=\frac{2\pi R \sin^2\theta}{g_5^2} \bigg[\gamma_E+\psi\left(1+\frac{(g_5Rf)^2}{2k_s}\right)
 +\frac{(g_5Rf)^2}{2k_s}\psi_1\left(1+\frac{(g_5Rf)^2}{2k_s}\right)\bigg].
  \ee
  
For a more intuitive understanding of $\Pi_{LR}(q^2)$ we also provide it in terms of
the resonance decomposition
\begin{align}\label{LR1}
\Pi_{LR}(q^2)&= q^4 \sin^2\theta\left(\sum\limits_n\frac{F^2_A(n)}{M^2_A(n)(q^2-M^2_A(n))}-\sum\limits_n\frac{F^2_V(n)}{M^2_V(n)(q^2-M^2_V(n))}\right)+
\\& + q^2\sin^2\theta\left(\sum\limits_n\frac{F^2_V(n)}{M^2_V(n)}-\sum\limits_n\frac{F^2_A(n)}{M^2_A(n)}\right)+F^2\sin^2\theta\notag.
\end{align}
Rearranging slightly the terms we may write it in resemblance to a well-known result for the contribution of
composite particles to the left-right part of the EW effective action~\cite{PT,Pich_2013}:
\be
\mathcal L_{eff}\supset\frac{gg'}4\sin^2\theta{W}_\mu^{\alpha}{B}_\nu^{\alpha}\left(q^\mu q^\nu-q^2\eta^{\mu\nu}\right)
\left(\frac{F^2}{q^2}+\sum\limits_n\frac{F^2_A(n)}{q^2-M^2_A(n)}-\sum\limits_n\frac{F^2_V(n)}{q^2-M^2_V(n)}\right).
\ee
Eventually, the $S$ parameter in terms of the masses and decay constants of the vector composite states
gets a form
\be \label{Spar2}
S=4\pi\sin^2\theta \bigg[\sum\limits_n\frac{F^2_V(n)}{M^2_V(n)}-\sum\limits_n\frac{F^2_A(n)}{M^2_A(n)}\bigg].
\ee
Recall that in our description $F_V(n)=F_V$ for all values of $n$.

Let us now investigate the validity of the equivalent of the Weinberg sum rules that
relate  the imaginary part of $\Pi_{LR}(q^2)$ to the masses and decay constants of the vector resonances
in the broken and unbroken channels. 
The way to proceed is to equate $\Pi_{unbr}$ to its subtracted counterpart given by
Eqn. (\ref{FV}) and do the same with the equivalent expressions in the broken sector ({\it i.e.} take Eqn.~(\ref{brA6})). One selects a suitable integration
circuit and formally 
\begin{align}\label{Vint}
\frac1\pi\int_0^\infty \frac{dt}{t}  \text{Im} \Pi_{unbr}(t)=\sum\limits_nF_V^2(n),\\ \label{Aint}
\frac1\pi\int_0^\infty \frac{dt}{t}  \text{Im} \Pi_{br}(t)=\sum\limits_n F_A^2(n)+F^2.
\end{align}
However, these expressions are ill-defined as neither the imaginary part of the poles has been properly
defined (in the resonance expansions (\ref{FV}) and (\ref{brA6})) nor does the external contour vanish. Clearly, the left hand sides
 of the above expressions are generically divergent.
In addition, the sum over resonances should possess an essential singularity on the real axis when the number of 
resonances $N_{max}$ encircled in the contour tends to infinity. 

In order to define the sum over the resonances more correctly 
we introduce the imaginary parts proportional to the masses following Vainshtein, {\it i.e.} we replace $M_V^2 (n)$ in
Eqn.~(\ref{FV}) to $M_V^2 (n)(1 - i\epsilon)$. This prescription reproduces the correct residues.

As is well known the convergence properties of the integrals on the 
left hand side of
(\ref{Vint}) and (\ref{Aint}) are greatly improved if
one considers $\Pi_{LR}(q^2)=\sin^2\theta \left(\Pi_{unbr}(q^2)-\Pi_{br}(q^2)\right)$. Let us
introduce for uniformity the sum  $F^2=\sum\limits_{n<N_{max}}F^2(n)$ (see Eqn. (\ref{brA6})) and
therefore consider 
\be\label{WSR1}
\frac{1}{\pi}\int_0^{M^2(N_{max})} \frac{dt}{t} \text{Im} \Pi_{LR}(t)
=\sin^2\theta\sum\limits_{n<N_{max}}(F_V^2(n) - F_A^2(n)- F^2(n)).
\ee
In QCD this integral vanishes because $\Pi_{LR}$ decays fast enough
to make the external contour contribution negligible
if $N_{max}$ is large enough. 
The equality of (\ref{WSR1}) to zero is the first Weinberg sum rule. 
The same argument works as well for the second sum rule to hold:
\be\label{WSR2}
\frac{1}{\pi}\int_0^{M^2(N_{max})} dt \text{Im} \Pi_{LR}(t)
=\sin^2\theta\sum\limits_{n<N_{max}}(F_V^2(n)M_V^2(n) - F_A^2(n)M_A^2(n))=0.
\ee
In fact, in QCD one gets a fairly good agreement with phenomenology by just including the first resonances~\cite{Weinberg67}.
In any case, the fact that the dispersion relation is convergent (no subtraction needed) indicates
that the limit $N_{max}\to \infty$ could be taken.

We would like to understand the analogous situation in the present theory. Then, the two questions arise: (a) can the
contour integral be neglected? (b) If this is the case, is the integral over the imaginary part 
along the real axis converging?

To answer the first question we derive the large $Q^2$ expansion of $\Pi_{LR}(Q^2)/Q^2$. In order to do that
we use again the Stirling's expansion of the $\psi$ function in Eqn.~(\ref{LR0}) and get for 
the left-right correlator
\be
\frac{\Pi_{LR}(Q^2)}{Q^2}=\sin^2\theta\frac{\kappa^2(fR)^2}{Q^2}\frac R{k_s}
\left(\ln \frac{Q^2}{4\kappa^2}+\ln\kappa^2\varepsilon^2+1+2\gamma_E-
\frac{g_5^2}{k_s}\frac{\kappa^2(fR)^2}{Q^2}\right)+\mathcal O\left(\frac1{Q^6}\right).
\ee
This limit is valid only in the (unphysical) region of $|\arg Q^2|<\pi$. The value on the
physical axis ($0 < q^2=-Q^2$) is ill-defined (needs a prescription, such as the one discussed above). However, to discuss the
convergence of the outer part of the circuit in order to be able to derive Eqns. (\ref{WSR1}) and
(\ref{WSR2}) this is all we need. Unlike in the QCD case the correlator does not vanish fast enough
due to the presence of the $\ln Q^2/Q^2 $ and $1/Q^2$ terms. Therefore the corresponding
dispersion relation requires one subtraction 
\be
\frac{\Pi_{LR}(Q^2)}{Q^2}=\int_0^\infty\frac{dt}{t+Q^2-i\epsilon} \frac1\pi\frac{\text{Im} \Pi(t)}{t}+a,
\ee 
$a$ being the subtraction constant,{\it i.e.} the part of $\Pi_{LR}(Q^2)$ not determined by 
its imaginary part.

In the deep Euclidean region one could use an expansion
\be\label{LR2}
\frac{1}{t+Q^2}=\frac1{Q^2}-\frac1{Q^2}t\frac1{Q^2}+...\ee
and then
\be \label{LR3}
\frac{\Pi_{LR}(Q^2)}{Q^2}=a+ \frac1{Q^2}\frac1\pi\int_0^\infty \frac{dt}{t}  \text{Im}\Pi_{LR}(t)-\frac1{Q^4}\frac1\pi\int_0^\infty dt 
 \text{Im}\Pi_{LR}(t)+ \ldots 
\ee

Let us now consider a large, but finite, number of resonances in the analysis. That is, $N_{max} <\infty$, {\it i.e.} the theory
is endowed with a cut-off that has to be connected to $N_{max}$ via the relation
$\ln N_{max}= -2\gamma_E -\ln\kappa^2\varepsilon^2$, as we saw before. The dispersion relation still holds and a value for the
constant $a$ (that depends logarithmically on the cut-off) is
provided. Then, the expansion (\ref{LR3}) can be compared order by order with
the large $Q^2$ expansion given in Appendix~\ref{Q2exp}, derived assuming a finite number of resonances.
Since there is no $1/Q^2$ term in the expansion (\ref{largeQ2}) of $\Pi_{LR}(Q^2)/Q^2$ 
we have
\be 
\int_0^{M^2(N_{max})} \frac{dt}{t}  \text{Im}\Pi_{LR}(t) =0, 
\ee
and consequently that $ \sum\limits_{n<N_{max}}(F_V^2(n) - F_A^2(n)-F^2(n)) = 0$. This establishes the formal validity of the
first Weinberg sum rule, provided that a finite number of resonances is kept.

However, the situation is very different to the one in QCD. Here $F^2$ is actually logarithmically dependent
on the cut-off (recall Eqn. (\ref{brA2})). This proves that the sum over vector resonances 
$\sum\limits_{n<N_{max}}(F_V^2(n) - F_A^2(n))$
is itself cut-off dependent if $N_{max} \to \infty$ implying that symmetry restoration takes place very slowly 
in the ultraviolet. 

Finally, regarding the second term in the expansion (\ref{LR3}) -- the absence of the corresponding term
in the expansion (\ref{largeQ2}) leads to
\be
\frac1\pi\int_0^{M^2(N_{max})} dt \text{Im}\Pi_{LR}(t)=0,
\ee
that proves the second Weinberg sum rule to be present in our model under the same assumptions
as for the first. Again, both the integral of the imaginary part over the real axis and
the sum over resonances are logarithmically divergent unless a cut-off is imposed.

\section{Applying the previous results to phenomenology}\label{phen}
In this section we provide a numerical estimate of the masses of new composite states.

First, let us review the spectra in all the channels and the model parameters appearing there.
We have degenerate vector and scalar states in the unbroken sector:
\be M_V^2(n)=M_\sigma^2(n)=4\kappa^2(n+1),\quad n=0,1,2...\ee

In the broken sector there are massless Goldstone bosons and their massive excitations, the
vectorial fields have a constant shift in the intercept relatively to $M_V^2(n)$:
\be M_A^2(n)=4\kappa^2\left(n+1+\frac{(g_5Rf)^2}{2k_s}\right),\quad
 M_\Pi^2(n)=4\kappa^2 n,\quad n=0,1,2...\ee

 The range of all the masses is defined by the $\kappa^2$ parameter originated in the dilaton ansatz
 of the soft wall holographic model. In $M_A$ we observe a dimensionless combination of the parameters 
 $fR$, where $f\neq0$ is responsible for the dynamical symmetry breaking $SO(5)\rightarrow SO(4)$.
 The parameters $g_5$ and $k_s$ are the free parameters of the 5D model fixed by a short-distance expansion
 of the two-point functions of an assumed fundamental theory and given in terms of $R$ and $N_{tc}$ 
 (see Appendix~\ref{loops}): 
 $\frac{g_5^2}{R}=\frac{4 \pi^2}{5N_{tc}},\ \frac{k_s}{R}=\frac{64\pi^2}{5N_{tc}}$,
 resulting in $\frac{g_5^2}{2k_s}=\frac{1}{32}$ and therefore $M_A^2(n)=4\kappa^2(n+1+
\frac{(fR)^2}{32})$.

To have a contribution of the new physics at a realistic degree 
we consider the bounds on the $S$ parameter,
calculated using 5D techniques in Eqn.~(\ref{Spar1}) or (\ref{Spar2}). 
From~\cite{Gfitter_2014} we take 
\be-0.06\leq S\leq0.16.\label{ph1}\ee
The PDG~\cite{PDG2016} gives a slightly more constraining $-0.05\leq S\leq0.15$, or assuming another 
oblique parameter $U=0$: 
$-0.01\leq S\leq0.15$ (at $90\%$ CL).

\begin{figure}
	\includegraphics[scale=0.27]{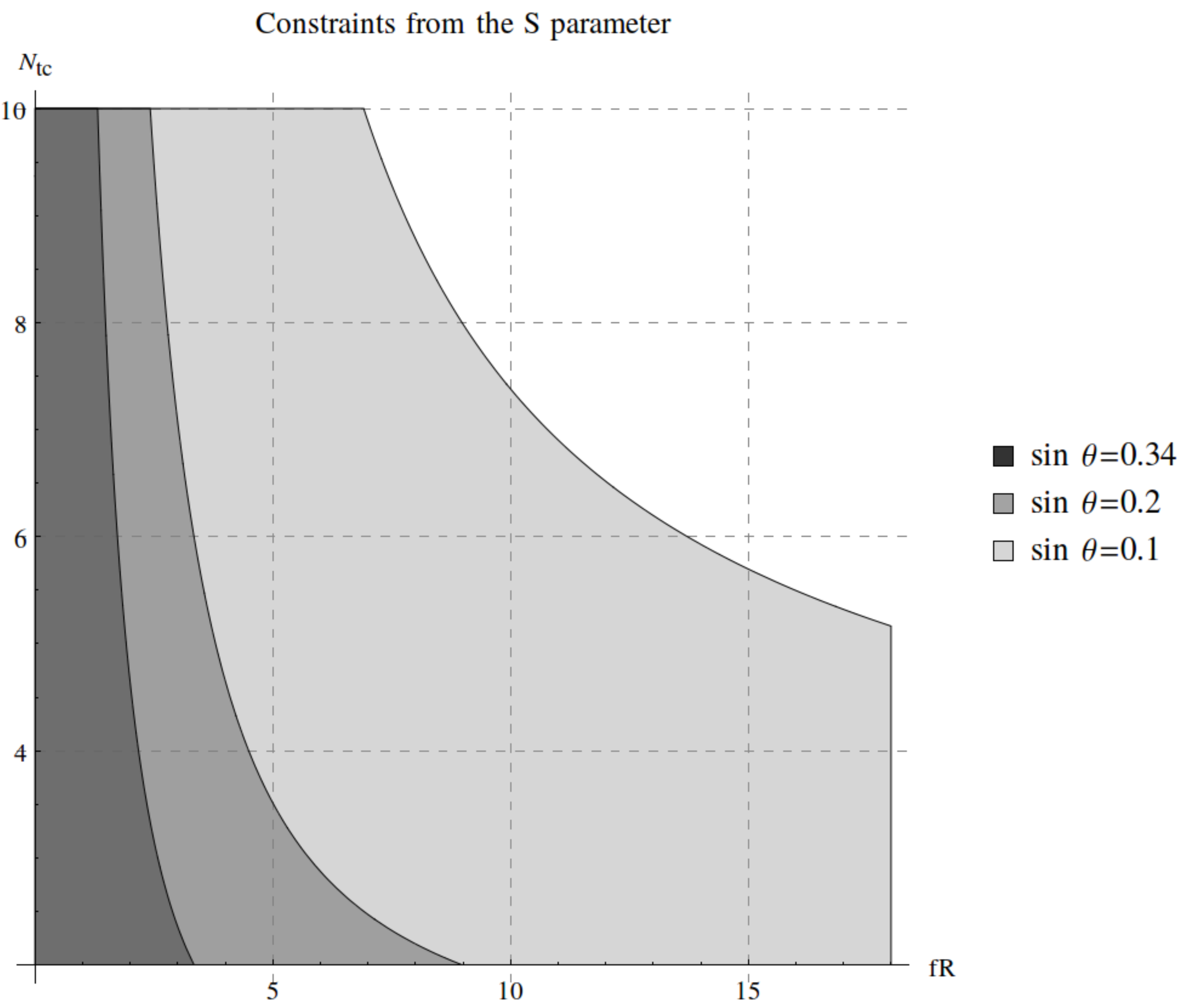}
	\caption{\label{Spar_fig} The $(\sin\theta, fR, N_{tc})$ parameter region allowed by the $S$ parameter restraints.}
\end{figure}

The $S$ parameter gives some restraint on a $(\sin\theta, fR, N_{tc})$ plane, see Fig.~\ref{Spar_fig}.
The larger the value of $\sin \theta$ the smaller the allowed region for $fR$ and $N_{tc}$, though 
we only consider $\sin\theta\leq0.34$ following the present bounds on the misalignment in
the Minimal Composite Higgs Model~\cite{Atlas_2015} (assuming the coupling of the Higgs to gauge bosons being $\kappa_V=\sqrt{1-\sin^2\theta}$).

We get no information about $\kappa^2$, the parameter that sets the overall scale of the masses, 
from the electroweak oblique parameters. However, through the $W$ mass of Eqn.~(\ref{MW1}) we
can relate it to the EWSB scale $v=246$ GeV.
Eqn.~(\ref{MW1}) as well contains the UV regulator 
\be\varepsilon=\frac1{\Lambda_\text{cut-off}}\simeq\frac1{4\pi F}=\frac{\sin\theta}{4\pi v}.\ee
The cut-off is assumed to be $4\pi F$ as this is the range of validity of the
effective theory of vector and scalar composite resonances
that has been assumed in the bottom-up holographic approach.
Thus, we get an implicit equation defining the $\kappa^2$ parameter:
\be\label{ph2}
\frac{v^2}{\sin^2\theta}+\frac{5}{256\pi^2}4\kappa^2N_{tc}(fR)^2\left(\ln\frac{4\kappa^2\sin^2\theta}{64\pi^2v^2}+2\gamma_E+\psi\left(1+\frac{(fR)^2}{32}\right)\right)=0.
\ee

\begin{table}[t]
 \centering
\caption{Different predictions of the minimal vector masses for $\sin\theta=0.2$ and $0.3$.}
\begin{tabular}{c | c | c | c | c | c}
$\sin\theta$ & $N_{tc}$ & $fR$  & $M_\ast=M_V(0)$, $\text{TeV}$ & $M_A(0)$, $\text{TeV}$ &$\sim N_{max}$\\
\hline
$0.2$& $2$ & $9.0$ & $1.02$ & $1.91$ & $290$ \\
$0.2$& $3$ & $5.8$ & $1.27$ & $1.81$ & $188$ \\
$0.2$& $4$ & $4.5$ & $1.39$ & $1.78$ & $156$ \\
$0.2$& $10$ & $2.4$ & $1.61$ & $1.75$ & $117$ \\
\hline
$0.3$& $2$ & $4.1$ & $1.61$ & $1.99$ & $51$ \\
$0.3$& $3$ & $3.1$ & $1.73$ & $1.97$ & $45$ \\
$0.3$& $4$ & $2.6$ & $1.78$ & $1.96$ & $42$ \\
$0.3$& $10$ & $1.5$ & $1.88$ & $1.95$ & $38$ \\
\end{tabular}
\label{tab-0}
\end{table}

\begin{figure}[t]
\centering
\subfigure{
\includegraphics[width=0.42\textwidth]{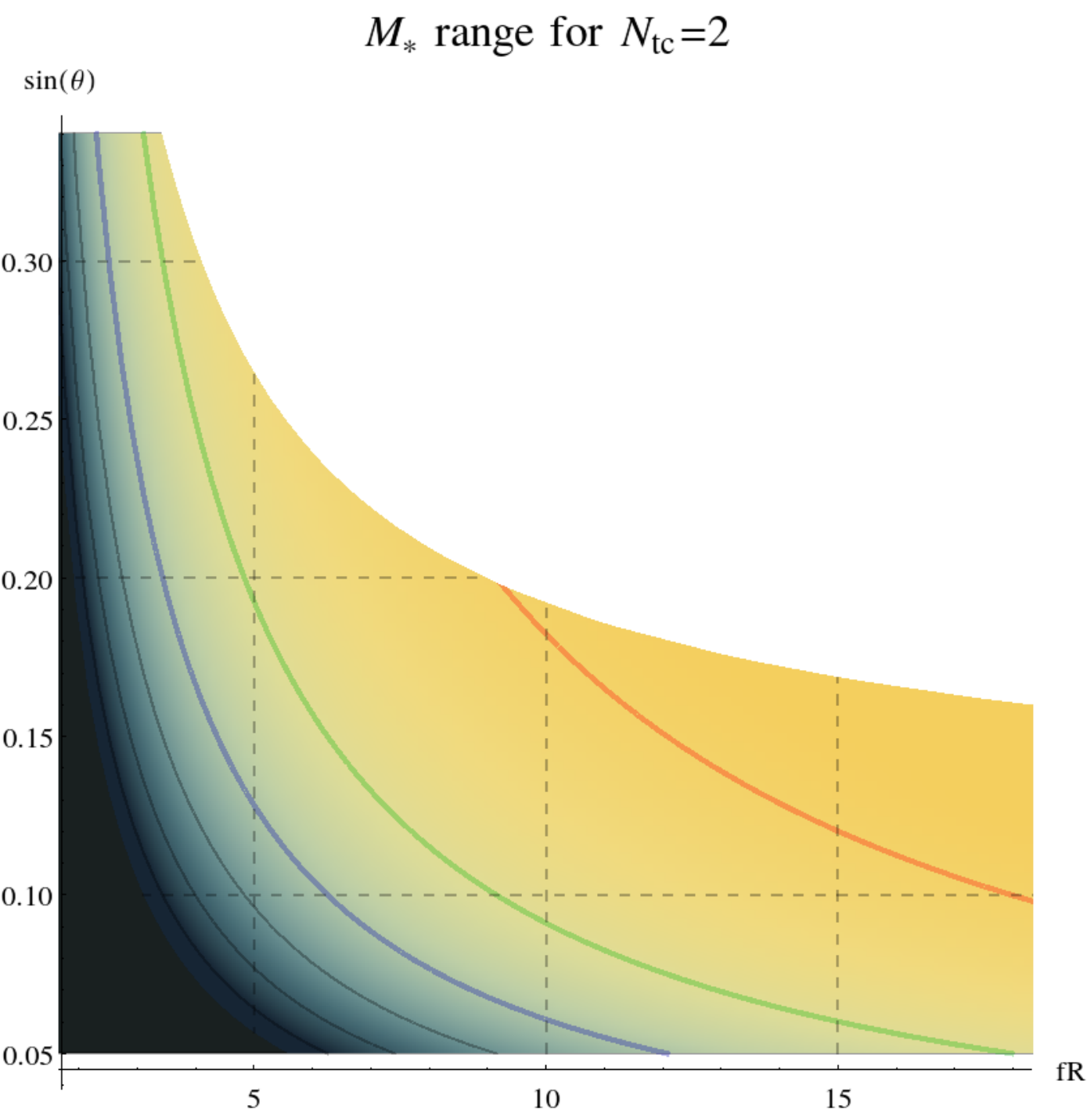}}
~
\subfigure{
\includegraphics[width=0.53\textwidth]{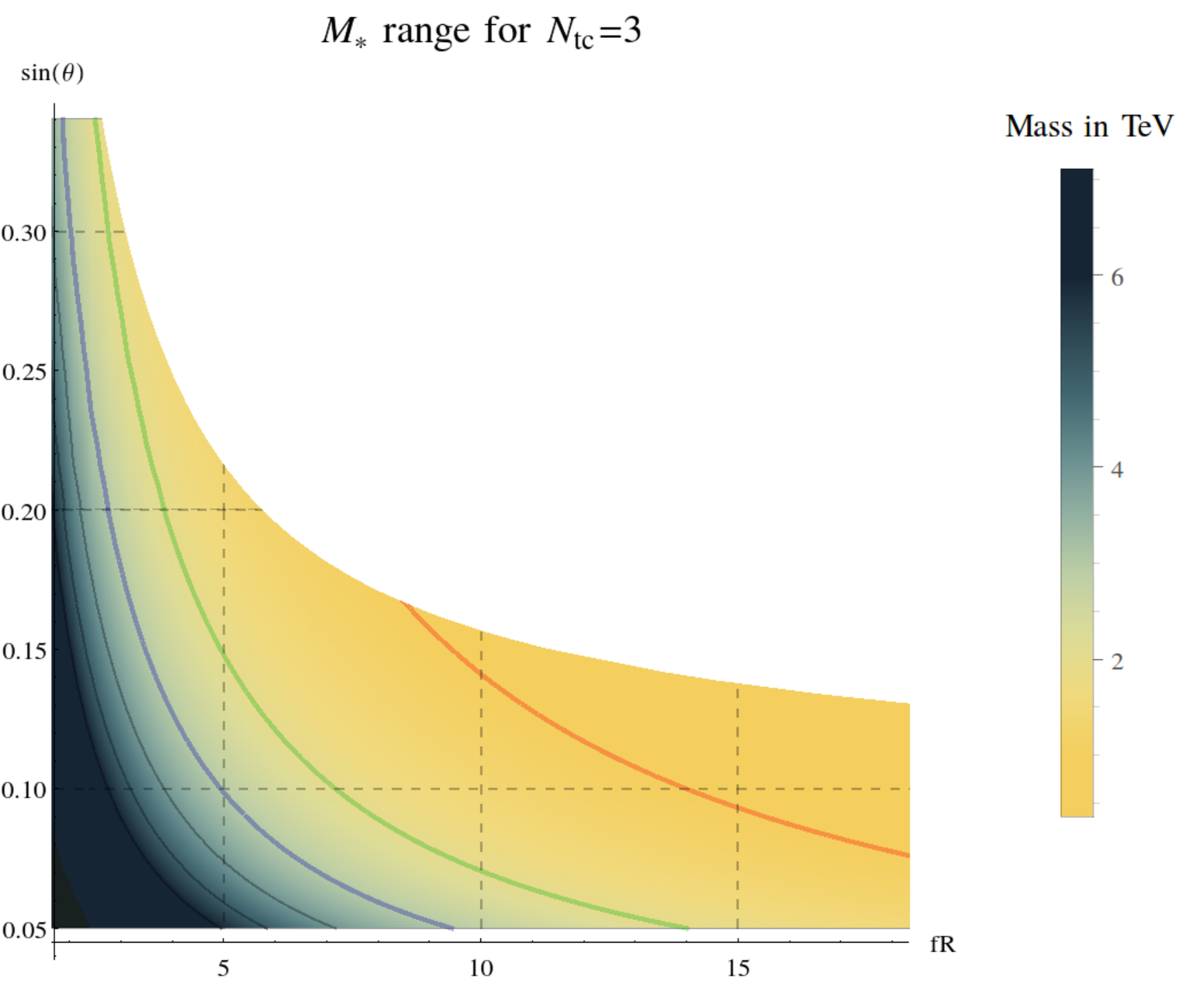}}

\subfigure{
\includegraphics[width=0.42\textwidth]{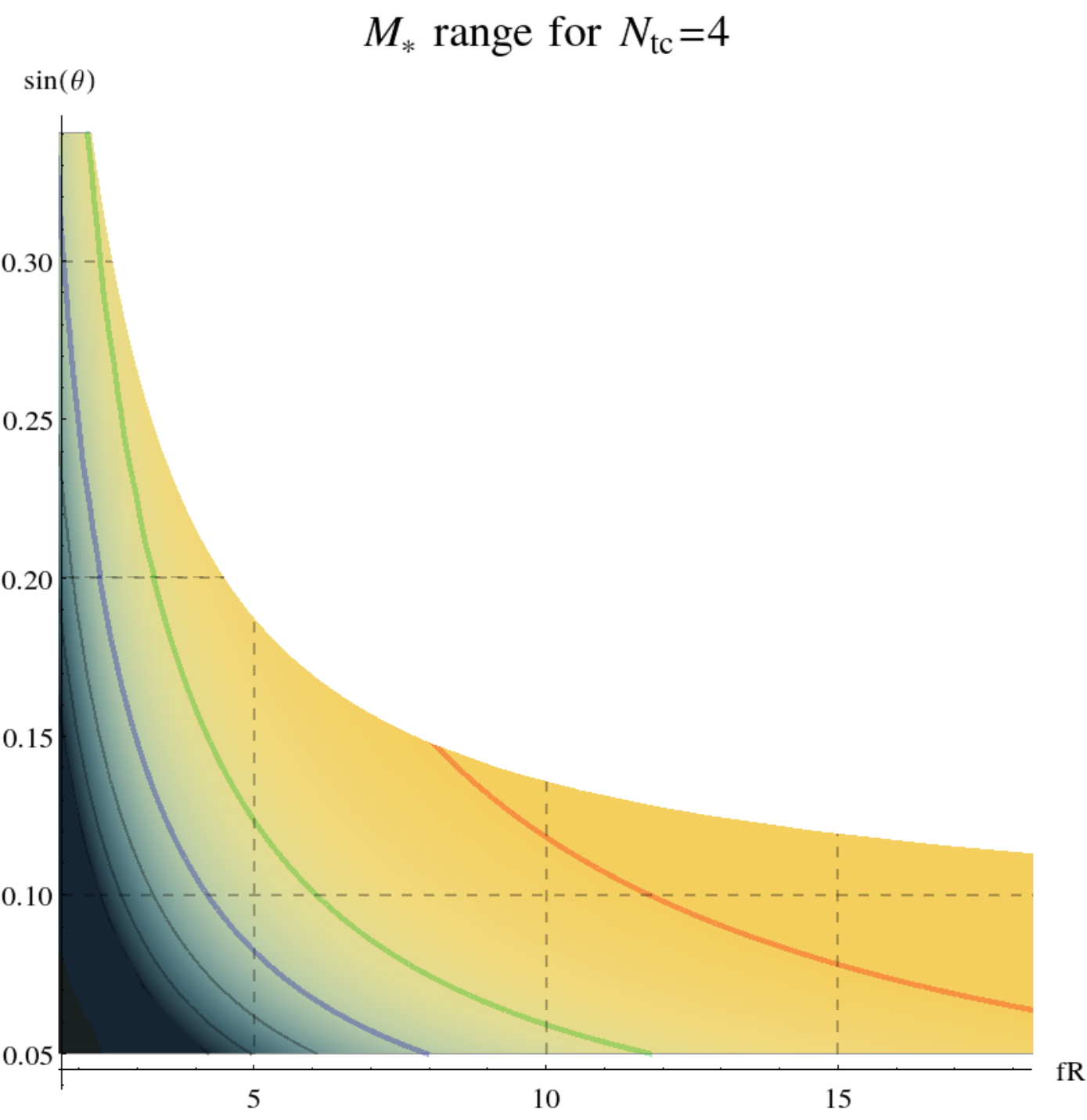}}
~
\subfigure{
\includegraphics[width=0.53\textwidth]{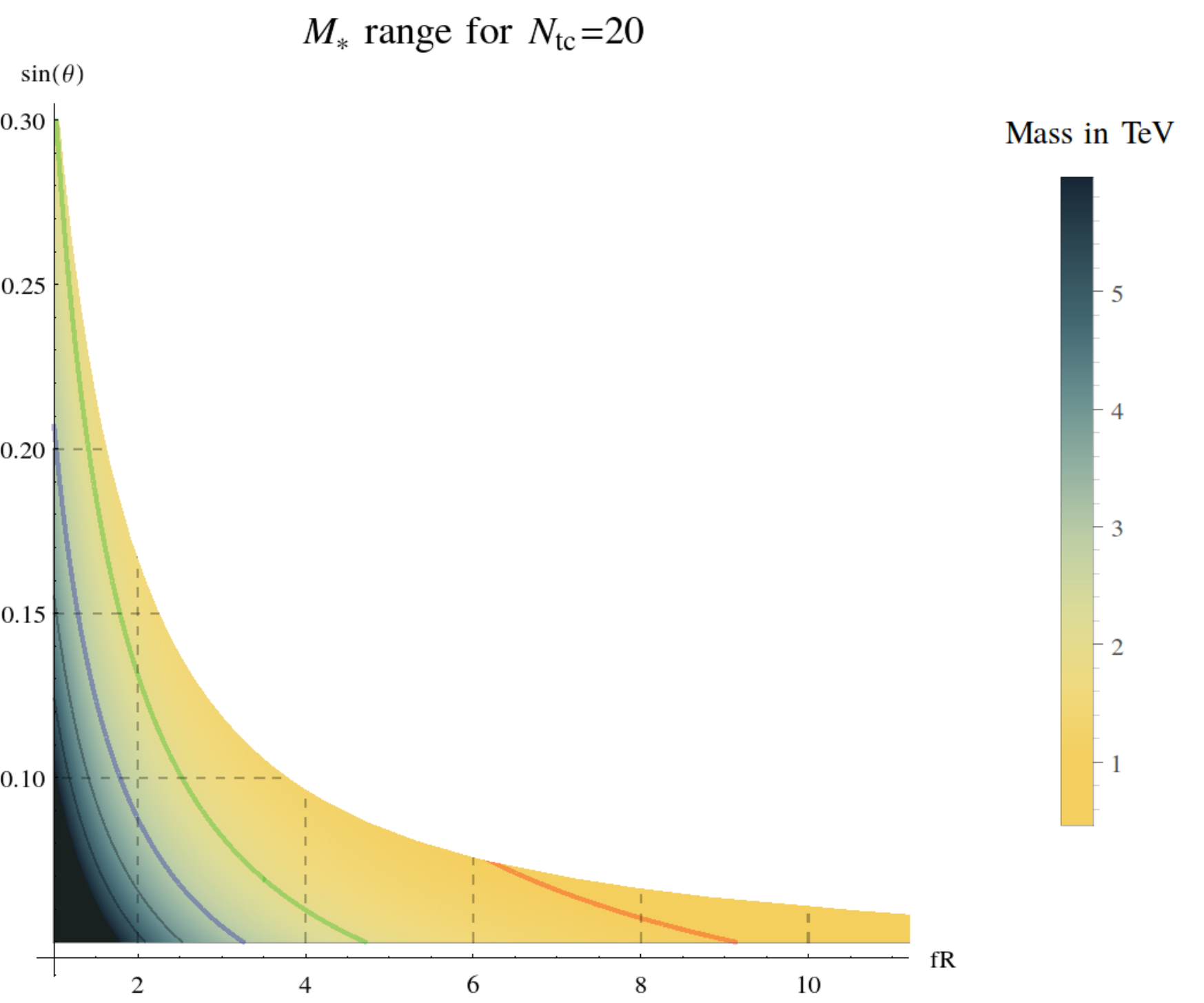}}

\caption{\label{Figmassrange} The density plots of $M_*$ for different values of $N_{tc}$.
The coloured curves represent the lines of constant $M_*$: the red one - $M_*=1$~TeV, 
the green one - $M_*=2$~TeV, the blue one - $M_*=3$~TeV and successive black curves for higher integer values. The white area 
represents the sector prohibited by the $S$ bound. }

\end{figure}

In addition, we have a connection between the UV cut-off $\varepsilon$ and the maximum number of resonances $N_{max}$
from which one can express $\kappa^2$:
\be
4\kappa^2=\frac{64\pi^2v^2}{\sin^2\theta}N_{max}^{-1}e^{-2\gamma_E}.
\ee

We consider Eqn.~(\ref{ph2}) as a problem of finding
a minimal value of a characteristic mass $M_\ast=\sqrt{4\kappa^2}$ depending on values
of the parameters $(\sin\theta, fR, N_{tc})$ 
from a region allowed by the $S$ parameter. We find that for fairly large values of 
$\sin\theta\simeq0.2 \div 0.34$ one 
can get $M_\ast$ of the order $\simeq 1 \div 2$~TeV and higher. See Table~\ref{tab-0}
for the lowest values of $M_V(0)$ and $M_A(0)$ in the allowed region of $(fR, N_{tc})$.
The lowest values of $M_\ast$ come from saturating the $S$ bound
with a value of $fR$ for a given fixed $N_{tc}$. The results presented in Table~\ref{tab-0}
correspond to the current maximum positive value of $S$. Should it be found that $S$ is 
$n$ times smaller, our estimates for $M_\ast$ become roughly $n$ times larger.

The predictions for the characteristic mass $M_*$ for a wide range of parameters are depicted 
at Fig.~\ref{Figmassrange}. First, we note that a broad variety of masses is allowed even 
considering the constraints of oblique corrections. The region below $1$~TeV starts for 
rather large values of $fR$ making them experimentally disfavoured unless the misalignment angle
is extremely small. In addition, a large $fR$ leads to a large splitting between vector fields
aligned in different (unbroken and broken) directions. 
Mind also that in a tower of resonances of one type we have a square root growth
with the number of a resonance, thus for a rather small value of $M_*$ we have a tower with several
low-lying states, like $M_V(n)=\{1, 1.4, 1.7, 2, ...\}$~TeV. 

Some general tendencies may be followed from Fig.~\ref{Figmassrange} as well. 
Consider the parameter space $(\sin\theta, fR, N_{tc})$ and fix any two values, then the growth 
of the third parameter results in lower $M_*$. Though unlimited growth in $fR$ results in 
unlikely small masses, the higher values of other two parameters soon face the upper 
experimental limit of the $S$ parameter.

We may imagine another free parameter of the theory considering a breaking pattern $SO(N) \rightarrow SO(N-1)$
with an arbitrary $N$.
A generalization from the particular case $N=5$ discussed previously is straightforward.
The minimality is lost of course, and many more states appear in the model. Still it is interesting to estimate 
how the value of $N$ affects the masses of the vector resonances.
Generally, the $S$ parameter has a linear dependence on $N$ and becomes proportionally more constraining at $N>5$, bringing the 
$S$ boundary to lower values of $fR$. That results in higher values for the characteristic mass $M_*$, but
the resonances of unbroken and broken sector are closer in mass.
For instance, having  $SO(9) \rightarrow SO(8)$ breaking pattern \cite{Bertuzzo2012} and $N_{tc}=3$ we 
may get the minimal values
$M_V(0)=1.49$~TeV and $M_A(0)=1.76$~TeV for $\sin\theta=0.2$ and  
$M_V(0)=1.83$~TeV and $M_A(0)=1.95$~TeV for $\sin\theta=0.3$.
However, $M_*$ cannot experience an uncontrollable growth,
the saturation is faced as soon as $fR$ is constrained by the $S$ to be small enough
to have the masses in two channels almost equal. Remarkably, though occurring for an unrealistically high degree of
the global group, these extreme values 
are rather moderate: $M_V(0)\simeq M_A(0)=1.7$~TeV for $\sin\theta=0.2$ 
and $M_V(0)\simeq M_A(0)=1.9$~TeV for $\sin\theta=0.3$, independently of the $N_{tc}$ value.

\section{Conclusions}
In this work we have reported on a bottom-up holographic study of the minimal composite Higgs model based
on the breaking pattern $SO(5)\to SO(4)$ and a gauge group misaligned with the unbroken group. The fundamental
degrees of freedom are assumed to be scalars living in some representation of $SO(5)$ and bound together by some unspecified
strong dynamics which is also assumed to trigger the breaking of the global symmetry. Extending our results
 to larger orthogonal groups would be straightforward. A possible extension in another direction would be
to consider Majorana fermions as fundamental degrees of freedom.

The main motivation for this analysis is to give plausible predictions for the spectrum of spin one 
and spin zero resonances taking into account all the existing experimental constraints at present. It has been
argued elsewhere that the $S$ parameter bounds force the lightest vector resonances to be at least in the 2 TeV region. 
Not much is known about possible scalar resonances so far.

The soft wall 5D holographic model we have used is inspired by the effective models of QCD and consists in a 
generalized sigma model coupled
both to the composite resonances and to the SM gauge bosons. The 5D model depends on three functions that parametrize
our ansatz: the dilaton $z$-profile (a feature common to all soft wall holographic models), and two functions $f(z)$ and
$b(z)$ that describe the generalized sigma model with an additional soft explicit breaking.
There are several interesting features present in the
resulting spectrum: one is that Goldstone bosons can be made exactly massless. Another one is that in the
unbroken sector vectors and scalars are degenerate in mass; not so for the states living in the broken
sector.

The two Weinberg sum rules hold but in a way only in a formal sense as the sum over resonances has to be cut off (it is
logarithmically divergent). The holographic effective theory has a built-in cut-off that is related to the maximum 
number of resonances included. However, adhering to this cut-off it is possible to derive relations involving resonance 
decay constants and masses. Yet, the fact that the sum rules are divergent implies that they are not saturated at all by
just the first resonance, as is the case in QCD.

We proceed to determining the minimal set of input parameters by including the short distance constraints resulting
from comparison with perturbation theory in the vector and scalar channels and include the constraints coming from
the $W$ mass, the $S$ parameter and the existing bounds on $\sin\theta$. This allows us to derive masses for the first
composite resonances. It is not difficult to find areas in the parameter space where a resonance between 1 and 2 TeV
is easily accommodated, even lighter in the lowest range of values for $\sin \theta$ and for large values of $fR$, even though
this limit looks somewhat unnatural and fine-tuned. Large values of $fR$ also lead to a large mass splitting between the
broken and unbroken sectors.

\begin{acknowledgments}
We are thankful to G. d'Ambrosio and D. Greynat for various early discussions on the subject. We would also 
like to thank S.S. Afonin
snd A.A. Andrianov for numerous conversations and, in the case of S.S.A. for a critical reading of the manuscript too.
We acknowledge financial support from the following grants:  FPA2013-46570-C2-1-P (MINECO), 2014SGR104 
(Generalitat de Catalunya), and MDM-2014-0369.

\end{acknowledgments}

\bibliography{biblio/composite_higgs_database}

\appendix
\section{Some properties of the confluent hypergeometric functions}\label{hyperf}
The confluent hypergeometric equation has the general form:
\be
x f''(x)+(c-x)f'(x)-af(x)=0.\ee

Solutions of this equation depend crucially on the value of the $a$ and $c$ parameters.
Here we provide a brief overlook of the properties of the confluent hypergeometric equation,
focusing on the dependence on the different integer values of the $c$ parameter~\cite{bateman1}.

For the positive integer values $c=1, 2, 3, ...$ we have 
\be f(x)=C_1\ _1F_1(a,c;x)+C_2 \Psi(a,c;x),\ee where $\ _1F_1(a,c;x)$ is called the Kummer's (confluent hypergeometric) function
and $\Psi(a,c;x)$ - the Tricomi's (confluent hypergeometric) function.

However, all the cases mentioned in the paper lie in the region of the non-positive 
integer $c$, for which one of the expected solutions, $_1F_1(a,c;x)$, does not exist, because it has poles at 
$c=0, -1, -2, ...$. In the same time the Tricomi's function can be analytically continued to any integer $c$.
Nevertheless, the fundamental system of solutions is rich enough and we are able to 
choose another two solutions:
\be\label{hyperfsol} f(x)=C_1 x^{1-c}\ _1F_1(a-c+1,2-c;x)+C_2 \Psi(a,c;x).\ee

Mark that the Tricomi's function exhibits a logarithmic behaviour for all integer $c$.
Specifically for the case $c=1-n,\ n=0,1,2,...$ one can write:
  \begin{align}\notag
  \Psi(a,1-n;x)&=\frac{(n-1)!}{\Gamma(a+n)}\sum\limits_{r=0}^{n-1}\frac{(a)_rx^r}{(1-n)_r r!}
  +\frac{(-1)^{n-1}}{n!\Gamma(a)}\bigg( \ _1F_1(a+n,n+1;x)x^n \ln x+\\ \label{Triclog}
  &+\left.\sum\limits_{r=0}^\infty  \frac{(a+n)_r}{(n+1)_r}[\psi(a+n+r)-\psi(1+r)-\psi(1+n+r)]\frac{x^{n+r}}{r!}\right),
  \end{align}
here the Pochhammer symbol is $(a)_n=1\cdot a\cdot(a+1)...(a+n-1)=\Gamma(a+n)/\Gamma(a)$,
$\psi(a)$ is the digamma function; and the first sum is absent for the case $n=0$.
There exists also a useful equation relating the Tricomi's functions of different arguments:
\be \label{Tricrel} \Psi(a,c;x)=x^{1-c}\Psi(a-c+1,2-c;x).\ee

The Kummer's function being an infinite series solution
$\ _1F_1(a,c;x)=\sum\limits_{n=0}^\infty \frac{(a)_n}{(c)_n}\frac{x^n}{n!}$
has a natural connection with the Laguerre polynomials (for integer $n>0,\ m>0$):
\be L_n^m(x)=\frac{(m+1)_n}{n!}\ _1F_1(-n,m+1,x).\ee

\section{Large $Q^2$ expansion of the correlator $\Pi_{LR}$}\label{Q2exp}

Here we perform the large $Q^2$ expansion of $\Pi_{LR}$  using the infinite series representation 
of the digamma function
$\psi(1+z)=-\gamma_E+\sum\limits_{n=1}^\infty \frac z{n(n+z)}$
valid for $z\neq-1,-2,...$ (derived from the series representation of the $\Gamma$-function)~\cite{bateman1}, for the particular ones from Eqn.~(\ref{LR0}) we have:
\begin{align}
&\lim\limits_{Q^2\rightarrow\infty}\psi\left(\frac{Q^2}{4\kappa^2}+1\right)=
-\gamma_E+\sum\limits_{n=0}^\infty\frac1{n+1}\sum\limits_{k=0}^\infty\left(\frac{-M^2_V(n)}{Q^2}\right)^k,\\
\notag&\lim\limits_{Q^2\rightarrow\infty}\psi\left(\frac{Q^2}{4\kappa^2}+1
+\frac{(g_5Rf)^2}{2k_s}\right)=-\gamma_E+\left(1+\frac{2\kappa^2(g_5Rf)^2}{k_sQ^2}\right)\sum\limits_{n=0}^\infty\frac1{n+1}
\sum\limits_{k=0}^\infty\left(\frac{-M^2_A(n)}{Q^2}\right)^k,
\end{align}
where for $k=0$ we have $\lim\limits_{N\rightarrow\infty}\sum\limits_{n=1}^N\frac1n
=\lim\limits_{N\rightarrow\infty}H_N=\ln N+\gamma_E+\mathcal O(1/N)$,
$H_N$ being the N-th harmonic number.

Substitution in Eqn.~(\ref{LR0}) yields order by order for $\Pi_{LR}(Q^2)/Q^2$:
\begin{align}
&\left(\frac1{Q^2}\right)^0:\quad \sin^2\theta\frac{R}{2g_5^2}\left(\sum\limits_{n=0}^\infty\frac1{n+1}-\sum\limits_{n=0}^\infty\frac1{n+1}\right);\\
&\left(\frac1{Q^2}\right)^1:\quad 4\kappa^2\sin^2\theta\frac{R}{2g_5^2}\sum\limits_{n=0}^\infty(1-1)-\sin^2\theta(fR)^2\kappa^2\frac{R}{k_s}\left(\ln\varepsilon^2\kappa^2+\gamma_E
+\sum\limits_{n=0}^\infty \frac1{n+1}\right);\\
&\left(\frac1{Q^2}\right)^2:\quad 4\kappa^4\sin^2\theta(fR)^2\frac{R}{k_s}\sum\limits_{n=0}^\infty(1-1).
\end{align}
Considering that $1$ and $-1$ come together for any $n$, as well as the fractions in the difference between harmonic sums,
we set these terms to zeros (certainly 0 for a finite sum) and get:
\be\label{largeQ2}
\frac{\Pi_{LR}(Q^2)}{Q^2}=-\sin^2\theta(fR)^2\kappa^2\frac{R}{k_s}\left(\ln\varepsilon^2\kappa^2+\gamma_E
+\sum\limits_{n=0}^\infty \frac1{n+1}
\right)\frac1{Q^2}+\mathcal O\left(\frac1{Q^6}\right).
\ee
In our discussion of subtraction constants in Section 
\ref{sec-corr} the term in brackets has been taken to be zero as the infinite sum is replaced with
the one up to $N_{max}$. Thus we show that the terms 
$1/Q^2$ and $1/Q^4$ are absent as long as $N_{max}<\infty$.

\section{Loop diagrams in the fundamental theory}\label{loops}
We presume that the fundamental theory is defined by an $SO(5)$-invariant Lagrangian:
\be\mathcal L _{str. int.} = \frac12\partial_\mu s_{\alpha\beta}\partial^\mu s^\top_{\beta\alpha}
-\frac12 m^2 s_{\alpha\beta}s^\top_{\beta\alpha}+\text{higher order terms},\ee
where $s_{\alpha\beta}$ are in a general 25-plet of $SO(5)$.
The propagator of $s_{\alpha\beta}$ is given then by:
\be
i\Delta^{\alpha\beta\alpha'\beta'}=
\frac {i\delta^{\alpha\alpha'}\delta^{\beta\beta'} }{p^2-m^2},
\ee
and the vertices are defined by the source-operator terms of Eqns. (\ref{Zqft}) and (\ref{scsource}). 
Given this we can compute the leading order diagrams for the scalar and vector two-point functions.

We begin with considering the scalar two-point function in terms of the fundamental fields:
\begin{align}\notag
 \int d^4xd^4y e^{-iq(x-y)} \langle i T^a_{\alpha\beta} s_{\beta\gamma}s_{\gamma\alpha}(x)\ \cdot\ 
i T^b_{\alpha'\beta'} s_{\beta'\gamma'}s_{\gamma'\alpha'}(y)\rangle
=\\
=i^2\left(-5N_{tc}\Tr T^aT^b\right) \int \frac{d^4k}{(2\pi)^4}\frac1{(k+q)^2-m^2}\frac1{k^2-m^2}.
\end{align}
As is mentioned in Section~\ref{sec-corr} in the large $Q^2$ limit the leading logarithmic
behaviour of the loop diagram corresponds to the one of $\Pi_S(Q^2)$:
$$\delta^{ab}\Pi_S(Q^2)\sim5\Tr T^aT^b\frac{N_{tc}}{16\pi^2}\ln Q^2,$$
giving the coefficient 
\be\label{ksfix}\frac{k_s}{R}=\frac{64 \pi^2}{5N_{tc}},\ee
where ``$N_{tc}$''
signifies the degree of the representation of the gauge group under which the
fundamental scalar fields transform (exactly $N_{tc}$ 
in case of the fundamental representation of $SU(N_{tc})$).

For the vector two-point function we get:
\begin{align}\notag
 \int d^4xd^4y e^{-iq(x-y)} \langle i[T^A,s]_{\alpha\beta}\partial^\mu s_{\alpha\beta} (x)\ \cdot\ 
i[T^B,s]_{\alpha'\beta'}\partial^\nu s_{\alpha'\beta'} (y)\rangle=\\
=\left(10N_{tc}\Tr T^AT^B\right) \int \frac{d^4k}{(2\pi)^4}\frac{ik^\mu iq^\nu}{((k+q)^2-m^2)(k^2-m^2)}. 
\end{align}

Taking the large $Q^2$ limit one may compute the leading logarithmic coefficient of the vectorial
two-point functions ($\Pi_{unbr}$ and $\Pi_{br}$). However, we do not get the full transverse structure
(only $q^\mu q^\nu$ term)
as the interaction considered does not come from a Lorentz-invariant term in the Lagrangian.
We miss here a coupling of a kind $ss A_\mu A^\mu$, which would have appeared had we 
considered real gauging in the fundamental theory. As this vertex could only contribute to the $\eta_{\mu\nu}$
part of $\Pi_{\mu\nu}$ and leaves $q^\mu q^\nu$ term unchanged we do not need any further computations
to fix the leading $\ln Q^2$ coefficient:
$$\delta^{AB}\Pi^{\mu\nu}(Q^2)\sim q^\mu q^\nu\left(-10\Tr T^AT^B\right)\frac{N_{tc}}{16\pi^2}\ln Q^2.$$
Thus, we get another coefficient fixed:
\be\label{g5fix}
\frac{g_5^2}{R}=\frac{4 \pi^2}{5N_{tc}}.\ee

\end{document}